\documentclass[preprint2]{aastex}

\shorttitle{Double Lobed Quasars}
\shortauthors{De Vries et al.}

\begin{document}
\title{Double Lobed Radio Quasars from the Sloan Digital Sky Survey}

\author{W. H. de Vries, R. H. Becker}
\affil{University of California, One Shields Ave, Davis, CA 95616}
\affil{Lawrence Livermore National Laboratory, L-413, Livermore, CA 94550}
\email{devries1@llnl.gov}

\and 

\author{R. L. White}
\affil{Space Telescope Science Institute, 3700 San Martin Drive,
Baltimore, MD 21218}

\begin{abstract}
  
We have combined a sample of 44\,984 quasars, selected from the Sloan Digital
Sky Survey (SDSS) Data Release 3, with the FIRST radio survey. Using a novel
technique where the optical quasar position is matched to the complete radio
{\it environment} within 450\arcsec, we are able to characterize the radio
morphological make-up of what is essentially an optically selected quasar
sample, regardless of whether the quasar (nucleus) itself has been detected in
the radio. About 10\% of the quasar population have radio cores brighter than
0.75 mJy at 1.4 GHz, and 1.7\% have double lobed FR2-like radio
morphologies. About 75\% of the FR2 sources have a radio core ($>0.75$mJy). A
significant fraction ($\sim40$\%) of the FR2 quasars are bent by more than 10
degrees, indicating either interactions of the radio plasma with the ICM or
IGM. We found no evidence for correlations with redshift among our FR2 quasars:
radio lobe flux densities and radio source diameters of the quasars have similar
distributions at low (mean 0.77) and high (mean 2.09) redshifts. Using a smaller
high reliability FR2 sample of 422 quasars and two comparison samples of
radio-quiet and non-FR2 radio-loud quasars, matched in their redshift
distributions, we constructed composite optical spectra from the SDSS
spectroscopic data. Based on these spectra we can conclude that the FR2 quasars
have stronger high-ionization emission lines compared to both the radio quiet
and non-FR2 radio loud sources. This is consistent with the notion that the
emission lines are brightened by ongoing shock ionization of ambient gas in the
quasar host as the radio source expands.

\end{abstract}


\section{Introduction}

Although we now know that the majority of quasars are, at best, weak radio
sources, quasars were first recognized as a result of their radio emission. Over
the decades a great deal of information has been accumulated about the radio
properties of quasars.  Generally speaking, roughly 10\% of quasars are thought
to be ``radio-loud'' \citep[e.g.,][and references therein]{kellermann89}. The
radio emission can be associated with either the quasar itself or with radio
lobes many kiloparsecs removed from the quasar (hereafter we refer these double
lobed sources as FR2s\footnote{Fanaroff \& Riley (1974) class II objects}).
Traditionally it was widely held that there was a dichotomy between the
radio-loud and radio-quiet quasar populations, although more recent radio
surveys have cast doubt on that picture
\citep[e.g.,][]{white00,cirasuolo03,cirasuolo05}. The advent of wide area radio
surveys like the FIRST survey coupled with large quasar surveys like SDSS permit
a more extensive inventory of the radio properties of quasars. The association
of radio flux with the quasar itself (hereafter referred to as core emission) is
straightforward given the astrometric accuracy of both the optical and radio
positions (typically better than 1 arcsec). The association of radio lobes is
more problematic since given the density of radio sources on the sky, random
radio sources will sometimes masquerade as associated radio lobes In this paper
we attempt to quantify both the core and FR2 radio emission associated with a
large sample of optically selected quasars.

Our new implementation of matching the FIRST radio {\it environment} to its
associated quasar goes beyond the simple one-to-one matching (within a certain
small radius, typically 2\arcsec), in that it investigates (and ranks) all the
possible radio source configurations near the quasar. This also goes beyond
other attempts to account for double lobed radio sources without a detected
radio core, most notably by \citet{ivezic02} who matched mid-points between
close pairs of radio components to the SDSS Early Data Release catalog.  While
this does recover most (if not all) of the FR2 systems that are perfectly
straight, it misses sources that are bent. Even slight bends in large systems
will offset the midpoint enough from the quasar position as to be a miss.

The paper is organized as follows. The first few sections (\S~\ref{intro1}
through \S~\ref{intro2}) describe the matching process of the radio and quasar
samples. The results (\S~\ref{results}) are separated in two parts: one based on
statistical inferences of the sample as a whole, and one based on an actual
sample of FR2 sources. These two are not necessarily the same. The former
section (\S~\ref{sampRes1} through \ref{redshiftdeps}) mainly deals with
occurrence levels of FR2's among quasars, the distribution of core components
among these FR2 quasars, and their redshift dependencies. All these results are
based on the detailed comparison between the actual and random samples. In other
words, it will tell us {\it how many} FR2 quasars there are among the total,
however, it does not tell us {\it which ones} are FR2.

This is addressed in the second part of \S~\ref{results}, which deals with an
{\it actual} sample of FR2 quasars (see \S~\ref{SampleSpecific} on how we select
these). This sample forms a representative subsample of the total number of FR2
quasars we infer to be present in the initial sample, and is used to construct
an optical composite spectrum of FR2 quasars. Section~\ref{compspectra} details
the results of the comparison to radio quiet and non-FR2 radio loud quasar
spectra.

\section{Optical Quasar Sample} \label{intro1}

Our quasar sample is based on the Sloan Digital Sky Survey (SDSS) Data Release 3
\citep[DR3, ][]{abazajian05} quasar list, as far as it overlaps with the FIRST
survey \citep{becker95}. This resulted in a sample of 44\,984 quasars.  In this
paper we focus on the radio population properties of optically selected quasars.

\section{Radio Catalog Matching}

The radio matching question is not a straightforward one. By just matching the
core positions, we are biasing against the fraction of radio quasars which have
weak, undetected, cores. Therefore, this section is separated in two parts, Core
Matching, and Environment Matching. The former is the straight quasar-radio
positional match to within a fixed radius (3\arcsec\ in our case), whereas the
latter actually takes the distribution of radio sources in the direct vicinity
of the quasar in account. This allows us to fully account for the double lobed
FR2 type quasars, whether they have detectable cores or not.

\subsection{Faint Core matches}\label{FaintCoreMatches}

\begin{figure}[t]
\epsscale{1.0}
\plotone{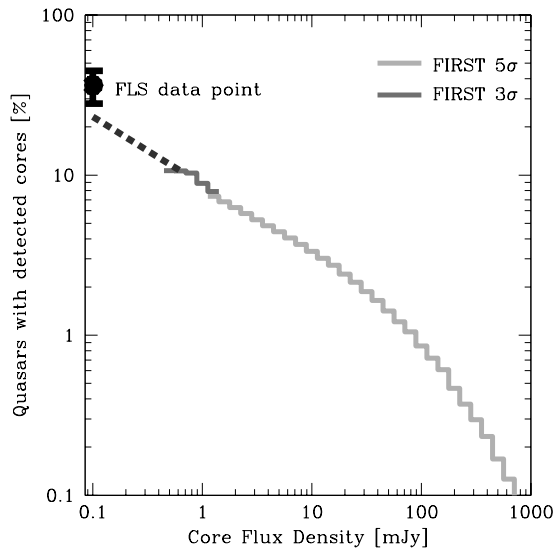}
\caption{Plot of the fraction of quasars with a detected radio core as function
of its flux density. The two lighter-grey curves are for the $5\sigma$ and
$3\sigma$ catalogs respectively. The dashed dark-grey line represents an
extrapolation of the expected number densities below our detection
threshold. The limiting detection rates are: 9.2\% ($5\sigma$), 11.8\%
($3\sigma$), and 23.3\% (extrapolation down to 0.1 mJy). The large dot
represents the detection rate for the Spitzer First Look Survey (FLS) field
($36\pm8$\%). Matching is done within 3 arcseconds.}
\label{corematches}
\end{figure}


In this section, we quantify the fraction of quasars that exhibit core emission.
We can actually go slightly deeper than the official FIRST catalog, with its
nominal $5\sigma$ lower threshold of 1.0 mJy, by creating $3\sigma$ lists based
on the radio images and quasar positions. This allows us to go down to a
detection limit of $\sim 0.75$ mJy (versus 1.0 mJy for the official version).

Given the steeply rising number density distribution of radio sources toward
lower flux levels, one might be concerned about the potential for an increase in
false detections at sub-mJy flux density levels. The relative optical to radio
astrometry is, however, accurate enough to match within small apertures (to
better than 3\arcsec), reducing the occurrence of chance superpositions
significantly. The surface density of radio sources at the 1 mJy level is not
high enough to significantly contaminate the counts based on 3 arcsecond
matching. The fraction of radio core detected quasars (RCDQ) out of the total
quasar population hovers around the 10\% mark, but this is a strong function of
the radio flux density limit. It also depends on the initial selection of the
quasar sample. The SDSS quasar selection is mainly done on their optical colors
\citep{richards02}, but $\sim3\%$ of the quasars were selected solely on the
basis of their radio emission (1397 out of 44\,984).  Looking at only those SDSS
quasars which have been selected expressly on their optical colors (see the
quasar selection flow-chart of Richards et al. 2002, Fig 1), there are 34\,147
sources which have either a QSO\_CAP, QSO\_SKIRT, and / or a QSO\_HIZ flag
set. For these, the relevant radio core detection fractions are 7.1\% (2430) and
10.1\% (3458) for the 5$\sigma$ and 3$\sigma$ detection limits, respectively
(the binomial error on these percentages is on the order of 0.1\%).  These core
fractions are higher for the 10\,837 quasars (44\,984$-$34\,147) that made it
into the SDSS sample via other means (1694, 15.6\% and 1855, 17.1\% for the 5
and 3$\sigma$ catalogs). The higher core fractions are due to the large number
of targeted FIRST sources that would not have made it into the sample otherwise,
and to the greater likelihood of radio emission among X-ray selected
quasars. Clearly, the initial quasar selection criteria impact the rate at which
their cores are detected by FIRST. The results have been summarized in Table~1.

A more direct view of the flux limit dependence of the RCDQ fraction is offered
by Fig.~\ref{corematches}. An extrapolation of the data suggests that at 0.1 mJy
about 20\%\ of quasar cores will be detected. This extrapolation is not
unrealistic, and even may be an underestimate: the extragalactic Spitzer First
Look Survey (FLS) field has been covered by both the SDSS and the VLA down to
0.1 mJy levels \citep{condon03}. Out of the 33 quasars in the DR3 that are
covered by this VLA survey, we recover 12 using the exact same matching
criterion. This corresponds to a fraction of 36\%, which carries an 8\% formal
$1\sigma$ uncertainty.

In fact, judging by the progression of detection rate in Fig.~\ref{corematches},
one does not have to go much deeper than 0.1 mJy to recover the majority of
(optically) selected quasars. The results and discussion presented in this
paper, however, are only relevant to the subset of quasars with cores brighter
than $\sim1$ mJy. It is this $\sim10\%$ of the total that is well-sampled by the
FIRST catalog.  This should be kept in mind as well for the sections where we
discuss radio quasar morphology.

\subsection{Environment Matching}\label{intro2}

\begin{figure*}[thb]
\epsscale{2.0}
\plotone{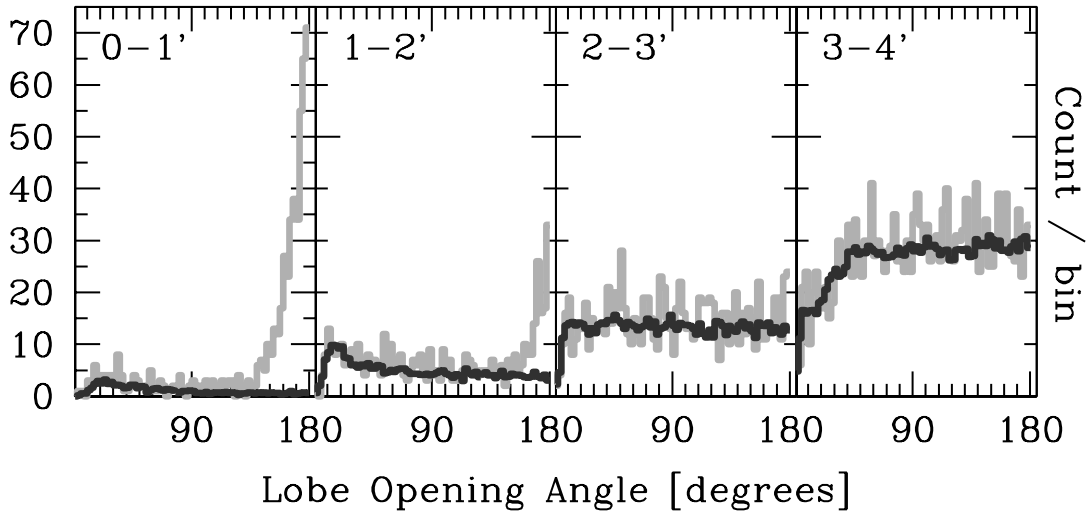}
\caption{Histograms of lobe opening angles of FR2 quasar candidates. Each box
represents a different FR2 size bin, as indicated by its diameter in
arcminutes. The light-grey histogram represents the candidate count, and in
dark-grey is the corresponding random-match baseline. This baseline increases
dramatically as one considers larger sources, while the FR2 candidate count
actually decreases.  Note both the strong trend toward linear systems (180
degrees opening angles), as well as the significant presence of {\it bent} FR2
sources. The bin size 2.5 degrees.}
\label{pahist}
\end{figure*}


The FIRST catalog is essentially a catalog of components, and not a list of
sources.  This means that sources which have discrete components, like the FR2
sources we are interested in, are going to have multiple entries in the FIRST
catalog. If one uses a positional matching scheme as described in the last
section, and then either visually or in an automated way assesses the quasar
morphology, one will find a mix of core- and lobe-dominated quasars {\it
provided} that the core has been detected.  However, this mechanism is going to
miss the FR2 sources without a detected core, thereby skewing the quasar radio
population toward the core dominated sources.

Preferably one would like to develop an objective procedure for picking out
candidate FR2 morphologies. We decided upon a catalog-based approach where the
FIRST catalog was used to find all sources within a 450\arcsec\ of a quasar
position (excluding the core emission itself). Sources around the quasar
position were then considered pairwise, where each pair was considered a
potential set of radio lobes. Pairs were ranked by their likelihood of forming
an FR2 based on their distances to the quasar and their opening angle as seen
from the quasar position. Higher scores were given to opening angles closer to
180 degrees, and to smaller distances from the quasar.  The most important
factor turned out to be the opening angle. Nearby pairs of sources unrelated to
the quasar will tend to have small opening angles as will a pair of sources
within the same radio lobe of the quasar, so we weighted against candidate FR2
sources with opening angles smaller than 50\arcdeg. The chances of these sources
to be real are very small, and even if they are a single source, their relevance
to FR2 sources will be questionable. We score the possible configurations as
follows:

\begin{equation}
w_{i,j} = \frac{\Psi / 50\arcdeg}{(r_i+r_j)^2}
\end{equation}

\noindent where $\Psi$ is the opening angle (in degrees), and $r_i$ and $r_j$
are the distance rank numbers of the components under consideration. The closest
component to the quasar has an $r=0$, the next closest is $r=1$, etcetera. This
way, the program will give the most weight to the radio components closest to
the quasar, irrespective of what that separation turns out to be in physical
terms. Each quasar which has at least 2 radio components within the 450\arcsec\
search radius will get an assigned ``most likely'' FR2 configuration (i.e., the
configuration with the highest score $w_{i,j}$. This, by itself, does not mean
it is a real FR2\footnote{Indeed, in some cases the most likely configurations
have either arm or lobe flux density ratios exceeding two orders of
magnitude. These were not considered further.}.

In fact, this procedure turns up large numbers of false positives. Therefore, as
a control, we searched for FR2 morphologies around a large sample of random sky
positions that fall within the area covered by FIRST. Since all of the results
for FR2s depend critically on the quality of the control sample, we increased
this random sample size 20-fold over the actual quasar sample (of
44\,984). Given the area of the FIRST survey ($\sim 9\,000$ sq. degree) and our
matching area (1/20th of a sq. degree), a much larger number of pointings would
start to overlap itself too much (i.e., the random samples would not be
independent of each other).

In Fig.~\ref{pahist} we display a set of histograms for particular FR2 sizes.
For each, the number FR2 candidates are plotted as a function of opening angle
both around the true quasar position (light-grey trace) as well as the offset
positions (dark-grey trace). There is a clear excess of nominal FR2 sources
surrounding quasar positions which we take as a true measure of the number of
quasars with associated FR2s.  Although the distribution of FR2s has a
pronounced peak at opening angles of 180 degrees, the distribution is quite
broad, extending out to nearly 90 degrees. It is possible that some of this
signal results from quasars living within (radio) clusters and hence being
surrounded by an excess of unrelated sources, but such a signal should not show
a strong preference for opening angles near 180 degrees.

The set of histograms also illustrates the relative importance of chance FR2
occurrences (dark-grey histograms), which become progressively more prevalent if one
starts considering the larger FR2 configurations. While the smallest size bin
does have some contamination ($\sim 14\%$ on average across all opening angles),
almost all of the signal beyond opening angles of 90 degrees is real (less than
5\% contamination for these angles). However, the significance of the FR2
matches drops significantly for the larger sources. More than 92\% of the signal
in the 3 to 4 arcminute bin is by chance. Clearly, most of the suggested FR2
configurations are spurious at larger diameter, and only deeper observations and
individual inspection of a candidate source can provide any certainty.

In the next few sections we describe the results of the analysis.

\section{Results}\label{results}

\subsection{Fraction of FR2 quasars}\label{sampRes1}

The primary result we can quantify is the fraction of quasars that can be
associated with a double lobed radio structure (whether a core has been detected
in the radio or not). This is different from the discussion in
\S~\ref{FaintCoreMatches} which relates to the fraction of quasars that have
radio emission at the quasar core position. This value, while considerably
higher than the rates for the FR2 quasars, does not form an upper limit to the
fraction of quasars associated with radio emission: some of the FR2 quasars do
not have a detected radio core.

\begin{figure}[t]
\epsscale{1.0} 
\plotone{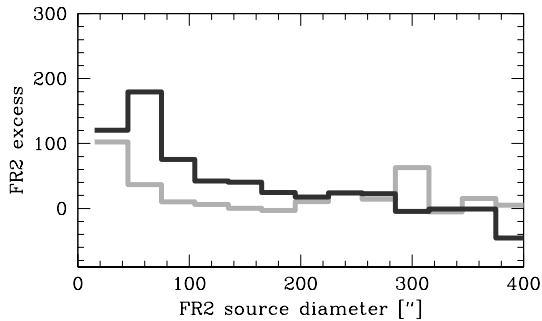}
\caption{Number of excess FR2 candidates over the random baseline numbers as
function of overall size. The histograms are for FR2 sources with cores
(dark-grey) and without cores (light-grey). The summed excess counts within
300\arcsec\ are 547 and 202 for the core and non-core subsamples,
respectively. Note that the smallest size bin for the core sample is affected by
resolution: it is hard to resolve a core inside a small double lobed structure.}
\label{candExcess}
\end{figure}


Figure~\ref{pahist} depicts the excess number of FR2 quasars over the baseline
values, plotted for progressively larger radio sources.  The contamination rates
go up as more random FIRST components fall within the covered area, and, at the
same time, fewer real FR2 sources are found. This effect is illustrated in
Fig.~\ref{candExcess}, which shows the FR2 excesses as function of overall
source size. The light-grey line indicates the FR2 number counts for candidates
{\it without} a detected (3$\sigma$) core, and the dark-grey histogram is for
the FR2 candidates {\it with} a detected core.  It is clear that FR2 sources
larger than about 300\arcsec\ are very rare, and basically cannot be identified
using this method. Most FR2 sources are small, with the bulk being having
diameters of less than 100\arcsec.

The summed total excess numbers, based on Fig.~\ref{candExcess} and limited to
300\arcsec\ or smaller, are 749 FR2 candidates (1.7\% of the total), of which
547 have cores. Some uncertainties in the exact numbers still remain,
particularly due to the noise in the counts at larger source sizes. A typical
uncertainty of $\sim 20$ should be considered on these numbers (based on
variations in the FR2 total using different instances of the random position
catalog).

At these levels, it is clear that the FR2 phenomenon is much less common than
quasar core emission; 1.7\% versus 10\% (see \S~\ref{FaintCoreMatches}).
Indeed, of all the quasars with a detected radio core, only about 1 in 9 is
also an FR2. The relative numbers have been recapitulated in Table~2.

\subsection{Core Fractions of FR2 quasars}

\begin{figure}[t]
\epsscale{1.0} 
\plotone{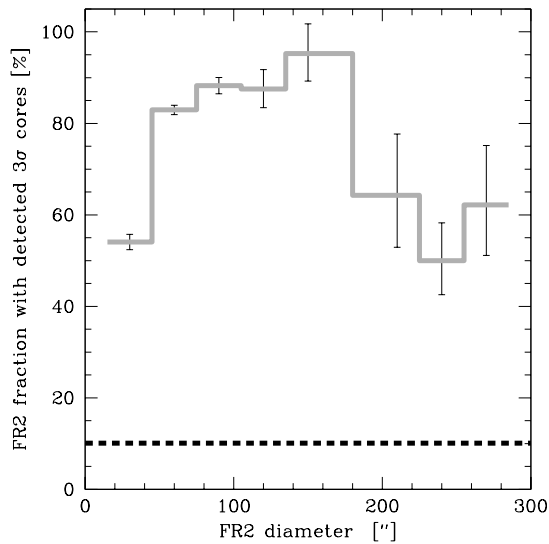}
\caption{Fraction of FR2 sources that have detected ($>0.75$ mJy) cores, as
function of overall size. As in Fig.~\ref{candExcess}, the smallest size bin is
affected by the angular resolution of FIRST.  The mean core fraction is 73.0\%,
which appears to be a representative value irrespective of the FR2 diameter The
horizontal dashed line represents the core-fraction of the non-FR2 quasar
population at 10.6\%.  The error estimates on the core fraction are a
combination of binomial and background noise errors.}
\label{coreFrac}
\end{figure}


As noted above, not all FR2 quasars have cores that are detected by FIRST.  We
estimate that about 75\% of FR2 sources have detected cores down to the 0.75 mJy
flux density level. This value compares well with the number for our ``actual''
FR2 quasar sample of \S~\ref{SampleSpecific}. Out of 422 FR2 quasar sources, 265
have detected cores (62.8\%) down to the FIRST detection limit (1 mJy).

We are now in the position of investigating whether there is a correlation
between the overall size of the FR2 and the presence of a radio core. In
orientation dependent unification schemes, a radio source observed close to the
radio jet axis will be both significantly foreshortened and its core brightness
will be enhanced by beaming effects (e.g., Barthel 1989, Hoekstra et
al. 1997). This would imply that, given a particular distribution of FR2 radio
source sizes and core luminosities, the smaller FR2 sources would be associated
(on average) with brighter core components. This should translate into a higher
fraction of detected cores among smaller FR2 quasars (everything else being
equal). Figure~\ref{coreFrac} shows the fraction of FR2 candidates that have
detected cores, as function of overall size. There does not appear to be a
significant trend toward lower core-fractions as one considers larger sources.
The much lower fraction for the very smallest size bin is due to the limited
resolution of the FIRST survey (about 5\arcsec), which makes it hard to isolate
the core from the lobe emission for sources with an overall size less than about
half an arcminute. Also, beyond about 275\arcsec\ the core-ratio becomes rather
hard to measure; not a lot of FR2 candidates are this large (see
Fig.~\ref{candExcess}).

Since the core-fraction is more or less constant, and does not depend on the
source diameter, it does not appear that relativistic beaming is affecting the
(faint) core counts. Unfortunately, one expects the strongest core beaming
contributions for the smallest sources; exactly the ones that are most affected
by our limited resolution.

\subsection{Bent Double Lobed Sources}

The angular distributions in Fig.~\ref{pahist} reveal a large number of more or
less {\it bent} FR2 sources. Bends in FR2 sources can be due to a variety of
mechanisms, either intrinsic or extrinsic to the host galaxy. Local density
gradients in the host system can account for bending \citep[e.g.,][]{allan84},
or radio jets can run into overdensities in the ambient medium, resulting in
disruption / deflection of the radio structure \citep[e.g.,][]{mantovani98}.
Extrinsic bending of the radio source can be achieved through interactions with
a (hot) intracluster medium. Any space motion of the source through this medium
will result in ram-pressure bending of the radio structure
\citep[e.g.,][]{soker88,sakelliou00}. And finally, radio morphologies can be
severely deformed by merger events \citep[e.g.,][]{gopalkrishna03}. Regardless
of the possible individual mechanisms, a large fraction of our FR2 quasars have
significant bending: only slightly more than 56\% of FR2 quasars smaller than 3
arcminutes have opening angles larger than 170 degrees (this value is 65\%\ for
the actual sample of \S~\ref{SampleSpecific}). This large fraction of bent
quasars is in agreement with earlier findings (based on smaller quasar samples)
of, e.g., \citet{barthel88, valtonen94, best95}.

\subsection{Redshift Correlations}\label{zdeps}

\begin{figure*}[th]
\epsscale{2.0} 
\plottwo{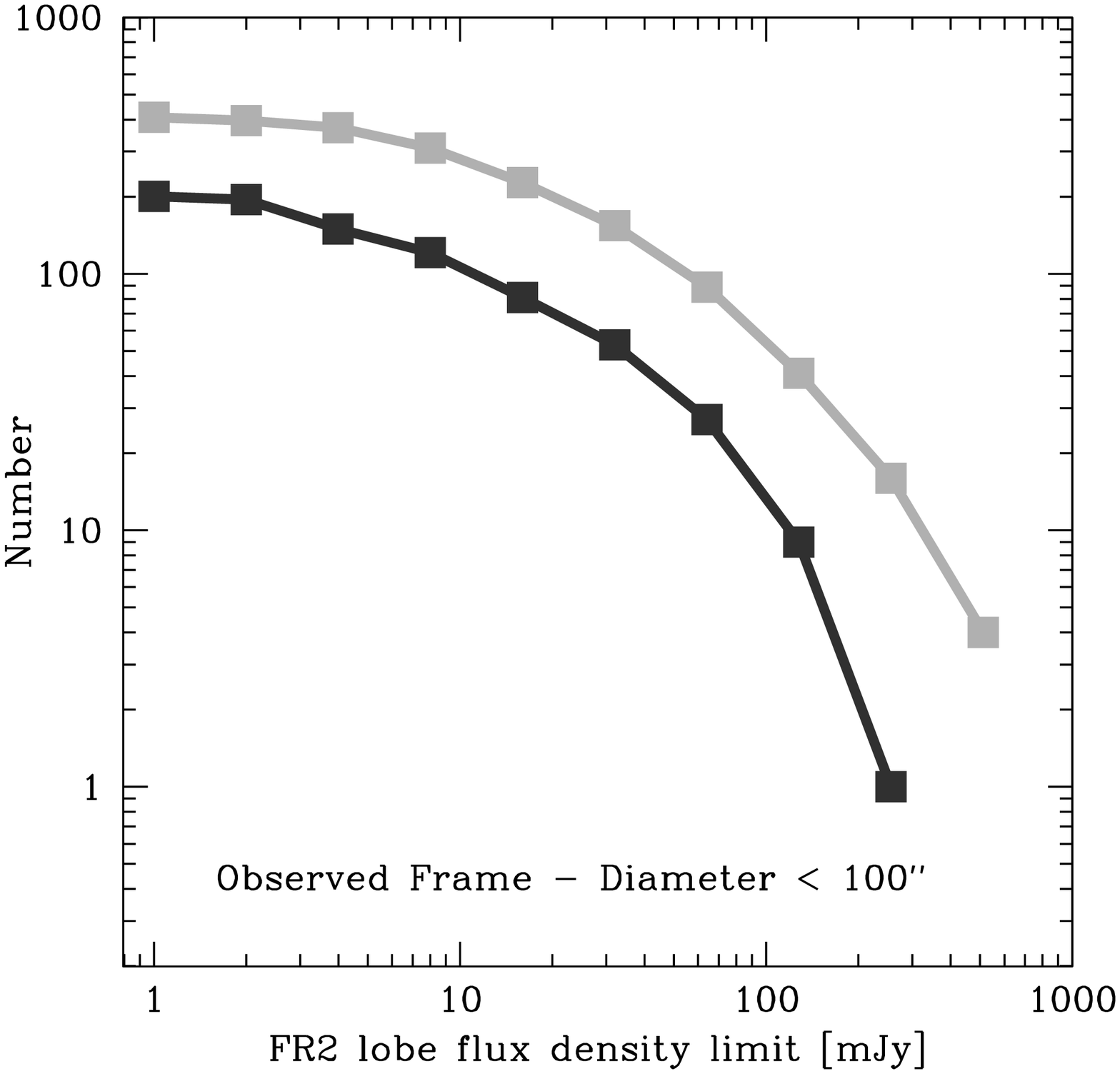}{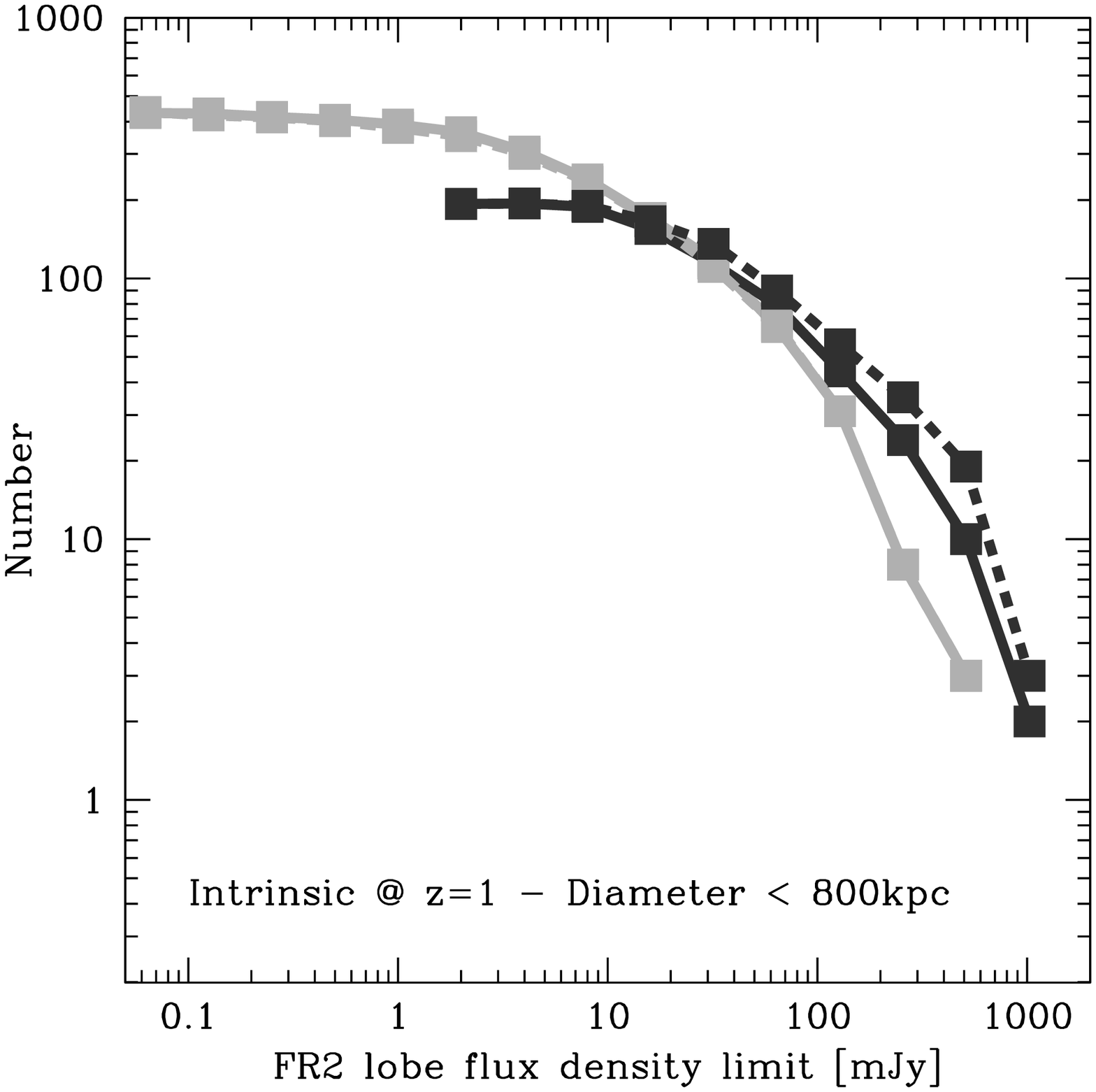}
\caption{Cumulative number of FR2 candidates as function of a lower threshold to
the lobe brightness. The left panel shows the distribution for the low-redshift
(light-grey line) and high-redshift (dark-grey) halves of the sample. The flux
limits are as observed (i.e., in mJy at 1.4GHz). There are about twice as many
low-redshift FR2 candidates as high-redshift candidates. In the panel on the
right the redshift dependencies have been taken out; all sources are placed on a
fiduciary redshift of 1 by k-correcting their lobe flux densities. Note that the
shape of the distribution only weakly depends on the assumed radio spectral
index used in the k-correction (from $\alpha=0$ to the canonical radio spectral
index of $\alpha=-0.75$, solid and dashed curves respectively).}
\label{zdep}
\end{figure*}


We can investigate whether there are trends with quasar redshift based on
statistical arguments.  This is done by subdividing the sample of 44\,984 in two
parts (high and low redshift), and then comparing the results for each
subsample. As with the main sample, each subsample has its own control sample
providing the accurate baseline.

Previous studies (e.g., Blundell et al. 1999) have suggested that FR2 sources
appear to be physically smaller at larger redshifts.  For self-similar expansion
the size of a radio source relates directly to its age. It also correlates with
its luminosity, since, based on relative number densities of symmetric double
lobed sources over a large range of sizes (e.g., Fanti et al. 1995, O'Dea \&
Baum 1997), one expects a significant decline in radio flux as a lobe expands
adiabatically.

While the picture in a fixed flux density limit will preferentially bias against
older (and therefore fainter) radio sources at higher redshifts, resulting in a
``youth-size-redshift'' degeneracy, scant hard evidence is available in the
literature.  Indeed, several studies contradict each other. See Blundell et
al. (1999) for a nice summary. We, however, are in a good position to address
this issue. First, our quasar sample has not been selected from the radio, and
as such has perhaps less radio bias built in. Also, we have complete redshift
information on our sample. The redshift range is furthermore much larger than in
any of the previous studies.  The median redshift for our sample of quasars is
1.3960, which results in mean redshifts for the low- and high-redshift halves of
the sample of $\overline{z_L}=0.769$, and $\overline{z_H}=2.088$.

The first test we can perform on the 2 subsamples is to check whether their
relative numbers as function of the average lobe flux density makes sense.  The
results are plotted in Fig.~\ref{zdep}, left panel. The curves are depicting the
cumulative number of FR2 candidates {\it smaller than 100\arcsec}\ for which the
mean lobe flux density is larger than a certain value. We have explicitly
removed any core contribution to this mean. We limited our comparison here to
the smaller sources, for which the background contamination is smallest.

Since the left panel shows the results in the observed frame, it is clear that
we detect far more local FR2 sources than high redshift ones (light-grey curve
in comparison to the dark-grey one). On average, more than twice as many
candidates fall in the low-redshift bin compared to the high-redshift ones (408
candidates versus 201). Furthermore, the offset between the two curves appears
to be fairly constant, indicating that the underlying population properties may
not be that different (i.e., we can match both by shifting the high-redshift
curve along the x-axis, thereby correcting for the lower lobe flux densities due
to their larger distances). This is exactly what we have done in the right
panel. All of the FR2 candidates have been put at a fiducial redshift of 1,
correcting its lobe emission both for relative distances and intrinsic radio
spectral index. For the former we assumed a WMAP cosmology (H$_{\rm o}=71$ km
s$^{-1}$ Mpc$^{-1}$, $\Omega_{\rm M} = 0.3$, and $\Omega_{\Lambda}=0.7$), and
for the latter we adopted $\alpha=-0.75$ for the high frequency part of the
radio spectrum ($> 1$ GHz).  It should be noted that these cumulative curves are
only weakly dependent on the cosmological parameters $\Omega$ and radio spectral
index $\alpha$. It is not dependent on the Hubble constant. The physical maximum
size for the FR2 candidates is set at 800kpc, which roughly corresponds to the
100\arcsec\ size limit in the left panel.

Both curves agree reasonably well now. At the faint end of each curve,
incompleteness of the FIRST survey flattens the distribution. This accounts for
the count mismatch between the light- and dark-grey curves below about 10
mJy. On the other end of both curves, low number statistics increase the
uncertainties.  The slightly larger number of bright FR2 sources for the
high-redshift bin (see Fig.~\ref{zdep}, right panel) may be real (i.e., FR2
sources are brighter at high redshifts compared to their low-redshift
counterparts), but the offset is not significant. Also note the effect of
changing the radio spectral index from $-0.75$ to $0$ (dotted dark-grey line
versus solid line). A negative $\alpha$ value has the effect of increasing the
lobe fluxes, especially for the higher redshift sources. Flattening the $\alpha$
to $0$ (or toward even more unrealistic positive values) therefore acts to lower
the average lobe fluxes, and as a consequence both cumulative distributions
start to agree better. This would also suggest that the high-redshift sources
may be intrinsically brighter, and that only by artificially lowering the fluxes
can both distributions made to agree.

\subsection{Physical FR2 sizes at low and high redshifts}\label{redshiftdeps}

\begin{figure}[tb]
\epsscale{1.0} 
\plotone{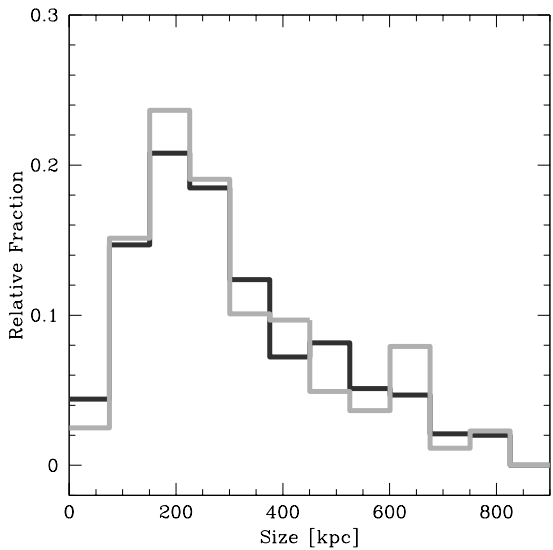}
\caption{Histogram of the FR2 source size distribution. The light-grey histogram
is for the low-redshift half of the sample, and the dark-grey line is for the
high-redshift sources. Both histograms have been corrected for random matches,
and therefore represent the real size distributions.}
\label{histSizes}
\end{figure}


This brings us to the second question regarding redshift dependencies: are the
high-redshift FR2 quasars intrinsically smaller because we are biased against
observing older, fainter, and larger radio sources?  To this end we used the
same two datasets that were used for Fig.~\ref{zdep}, right panel. The upper
size limit is set at 800kpc for both subsets, but this does not really affect
each size distribution, since there are not that many FR2 sources this
large. Figure~\ref{histSizes} shows both distributions for the low and high
redshift bins (same coloring as before). As in Fig.~\ref{coreFrac}, the smallest
size bins are affected by resolution effects, albeit it is easier to measure the
lobe separation than whether or not there is a core component between the lobes.
The smallest FR2 sources in our sample are about 10\arcsec\ ($\sim$80 kpc at
$z=1$), which is a bit better than the smallest FR2 for which a clear core
component can be detected ($\sim 30\arcsec$). The apparent peak in our
size distributions (around 200 kpc) agrees with values found for 3CR quasars.
\citet{best95} quote a value of $207\pm29$ kpc, though given the shape of the
distribution it is not clear how useful this measure is. 

A Kolmogorov-Smirnoff test deems the difference between the low-redshift (green
histogram) and high-redshift (blue histogram) size distributions
insignificant. Therefore, it does not appear that there is evidence for FR2
sources to be smaller in the earlier universe.

An issue that has been ignored so far is that we tacitly assumed that the FR2
sizes for these quasars are accurate. If these sources have orientations
preferentially toward our line of sight (and we are dealing with quasars here),
significant foreshortening may underestimate their real sizes by quite a bit
(see Antonucci 1993). This will also ``squash'' both distributions toward the
smaller sizes, making it hard to differentiate the two.

Previous studies \citep[e.g.,][]{blundell99} relied on (smaller) samples of
radio galaxies, for which the assumption that they are oriented in the plane of
the sky is less problematic. Other studies which mainly focused on FR2 quasars
(e.g, Nilsson et al. 1998, Teerikorpi 2001) also do not find a size-redshift
correlation.

\subsection{Sample Specific Results}\label{SampleSpecific}

The next few sections deal with properties {\it intrinsic} to FR2 quasars. As
such, we need a subsample of our quasar list that we feel consists of genuine
FR2 sources. We know that out of the total sample of 44\,984 about 750 are FR2
sources, however, we do not know which ones. What we can do is create a
subsample that is guaranteed to have less than 5\% of non-FR2 source
contamination. This is done by stepping through the multidimensional space
spanned up by histograms of the Fig.~\ref{pahist} type as function of overall
size. As can be seen in Fig.~\ref{pahist}, the bins with large opening angles
only have a small contamination fraction (in this case for sources smaller than
100\arcsec). Obviously, the signal-to-noise goes down quite a bit for larger
overall sizes, and progressively fewer of those candidates are real FR2. By
assembling all the quasars in those bins that have a contamination rate less
than 5\%, as function of opening angle and overall size, we constructed a FR2
quasar sample that is more reliable than 95\%\footnote{Actually, a visual
inspection of the radio structures yielded only 5 bogus FR2 candidates (i.e., a
98.8\% accuracy).}. It contains 422 sources, and forms the basis for our
subsequent studies. Sample properties are listed in Table~3, and the positions
of the 417 actual FR2 quasars are given in Table~4.

\subsection{FR2 Sample Asymmetries}

The different radio morphological properties of the FR2 sources have been used
with varying degrees of success to infer its physical properties. In particular,
these are: the observed asymmetries in the arm-length ratio ($Q$, here defined as
the ratio of the larger lobe-core to the smaller lobe-core separation), the lobe
flux density ratio ($F = F_{\rm lobe, distant} / F_{\rm lobe, close}$), and the
distribution of the lobe opening angle ($\Psi$, with a linear source having a
$\Psi$ of 180\arcdeg).

\citet{gopalkrishna04} provide a nice historic overview of the literature on
these parameters. As can be inferred from a median flux ratio value of $F < 1.0$
(see Table~5), the closer lobe is also the brightest. This is consistent with
the much earlier findings of \citet{mackay71} for the 3CR catalog, and implies
directly that the lobe advance speeds are not relativistic, and that most of the
arm-length and flux density asymmetries are intrinsic to the source (and not due
to orientation, relativistic motions, and Doppler boosting).

If we separate the low and high-redshift parts of our sample, we can test
whether any trend with redshift appears. \citet{barthel88}, for instance,
suggested that quasars are more bent at high redshifts. In our sample we do not
find a strong redshift dependency. The median opening angles are 173.6 and
172.7\arcdeg, for the low and high redshift bins respectively\footnote{Eqn.~1
does not bias against opening angles anywhere between 50\arcdeg and 180\arcdeg
(as indicated by the constant background signal in Fig.~\ref{pahist}), nor is it
dependent on redshift.}. A Kolmogorov-Smirnoff test deemed the two distributions
different at the 97.2\% confidence level (a 2.2$\sigma$ results). This would
marginally confirm the Barthel \& Miley claim. However, \citet{best95} quote a
2$\sigma$ result in the opposite sense, albeit using a much smaller sample (23
quasars).

We also found no significant differences between the low and high-redshift
values of the arm-length ratios $Q$ (KS-results: different at the 87.0\% level,
1.51$\sigma$), and the flux ratios $F$ (similar at the 97.0\% level,
2.2$\sigma$).

The Mackay-type asymmetry, in which the nearest lobe is also the brightest, is
not found to break down for the brightest of our quasars. If we separate our
sample into a low- and high-flux bin (which includes the core contribution), we
do not see a reversal in the flux asymmetry toward the most radio luminous FR2
sources \citep[e.g.,][and references therein]{gopalkrishna04}.  Actually, for
our sample we find a significant (3.25$\sigma$) trend for the brightest quasars
to adhere more to the Mackay-asymmetry than the fainter ones. 

\subsection{Control Samples}

Using the same matching technique as described in the previous section, we made
two additional control samples. Whereas our FR2 sample is selected based on a
combination of large opening angle ($\ga 150$ degrees) and small overall size
($\la 200$\arcsec), our control samples form the other extreme. Very few, if
any, genuine FR2 sources will be characterized by radio structures with small
opening angles ($< 100$ degrees) and large sizes ($> 450$\arcsec). Therefore, we
use these criteria to select two {\it non-FR2} control samples: one that has a
FIRST source coincident with the quasar (remember that the matching algorithm
explicitly excludes components within 3\arcsec\ of the quasar position), and
another one without a FIRST counterpart to the quasar. For all practical
purposes, we can consider the former sample to be quasars which are associated
with just one FIRST component (the ``core dominated'' sample - CD), and the
latter as quasars without any detected FIRST counterpart (the ``radio quiet''
sample - RQ).

Both of the CD and RQ samples initially contained more candidates than the FR2
sample. This allows for small adjustments in the mean sample properties, in
particular the redshift distribution. We therefore matched the redshift
distribution of the CD and RQ samples to the one of the FR2 sample. This resulted
in a CD sample which matches the FR2 in redshift-space and in absolute
number. The RQ sample, which will function as a baseline to both the FR2 and CD
samples, contains a much larger number (6330 entries), but again with an
identical redshift distribution. The mean properties of the samples are listed
in Table~3.

\subsection{Composite Optical Spectra}\label{compspectra}

\begin{figure}[tb]
\epsscale{1.0}
\plotone{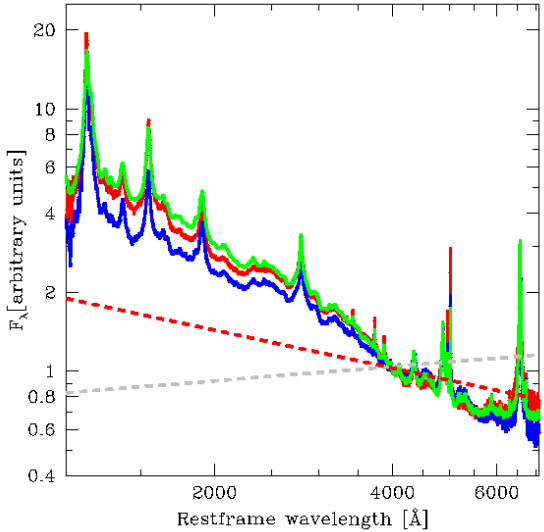}
\caption{Composite spectra for our three samples of quasars: radio quiet (RQ)
quasars in green, core dominated (CD) radio-loud quasars in blue, and lobe
dominated (FR2) radio-loud quasars in red. This plot can be directly compared to
Fig. 7 of Richards et al. (2003), and illustrates both the small relative color
range among our 3 samples (all fall within the ``normal'' range of Richards et
al.), and the apparent lack of significant intrinsic dust-reddening in these
quasars (the red and gray dashed lines represent moderate to severe levels of
dust-reddening). All spectra have been normalized to the continuum flux at
4000\AA.}
\label{compView}
\end{figure}

\begin{figure}[tb]
\epsscale{1.0} 
\plotone{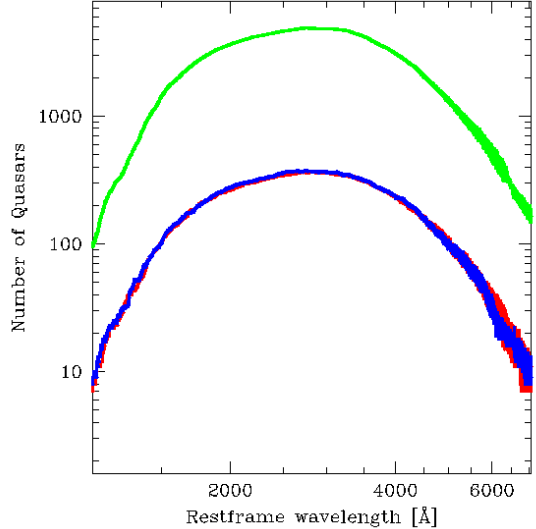}
\caption{Number of quasars that contributed to the composite spectrum as 
function of wavelength. The color-coding is the same as for Fig.~\ref{compView}.
The RQ sample (green histogram) contains 15 times as many quasars as both
the CD and FR2 samples.}
\label{nrSpec}
\end{figure}


One of the very useful aspects of our SDSS based quasar sample is the
availability of a large set of complementary data, including the optical spectra
for all quasars. An otherwise almost impossible stand-alone observing project
due to the combination of low FR2 quasar incidence rates and large datasets, is
sidestepped by using the rich SDSS data archive. We can therefore readily
construct composite optical spectra for our 3 samples (as listed in Table~3).
We basically used the same method for the construction of the composite spectrum
as outlined by \citet{vandenberk01}, in combination with a relative
normalization scheme similar to the ones used in \citet{richards03}. Each
composite has been normalized by its continuum flux at 4000\AA\ (restframe).

The resulting spectra are plotted in Fig.~\ref{compView}, color coded green for
the radio-quiet (RQ) quasar control sample, blue for the core-dominated (CD)
radio-loud quasars, and red for the lobe dominated (FR2) radio quasars. All
three composite spectra are similar to each other and to the composite of
\citet{vandenberk01}. Figure~\ref{nrSpec} shows the number of quasars from each
subsample that were used in constructing the composite spectrum. Since each
individual spectrum has to be corrected by a $(1+z)$ factor to bring it to its
restframe, not all quasars contribute to the same part of the composite.  In
fact, the quasars that contribute to the shortest wavelengths are not the same
that go into the longer wavelength part. This should be kept in mind if one
wants to compare the various emission lines. Any dependence of the emission
line properties on redshift will therefore affect the short wavelength part
of the composite more than the long wavelength part (which is made up of 
low redshift sources).

\citet{richards03} investigated the effect of dust-absorption on composite
quasar spectra (regardless of whether they are associated with radio sources),
and we have indicated two of the absorbed template spectra (composite numbers 5
and 6, see their Fig.~7) in our Fig.~\ref{compView} as the red and gray dashed
lines, respectively. From this it is clear that our 3 sub-samples do not appear
to have significant intrinsic dust-absorption associated with them. Indeed, the
range of relative fluxes toward the blue end of the spectrum falls within the
range of ``normal'' quasars (templates 1$-$4 of Richards et al. (2003)). The
differences in spectral slopes among our 3 samples are real. We measure
continuum slopes (over the range 1450 to 4040\AA, identical to Richards et
al. (2003)) of: $\alpha_{\nu}=-0.59\pm0.01$, $\alpha_{\nu}=-0.47\pm0.01$, and
$\alpha_{\nu}=-0.80\pm0.01$ for the FR2, RQ, and CD samples respectively. These
values are significantly different from the reddened templates
($\alpha_{\nu}=-1.51$ and $\alpha_{\nu}=-2.18$ for the red and grey dashed lines
in Fig.~\ref{compView}), suggesting that our quasars are intrinsically different
from dust-reddened quasars \citep[e.g.,][]{webster95,francis00}.

\begin{figure*}[p]
\epsscale{2.0} 
\plotone{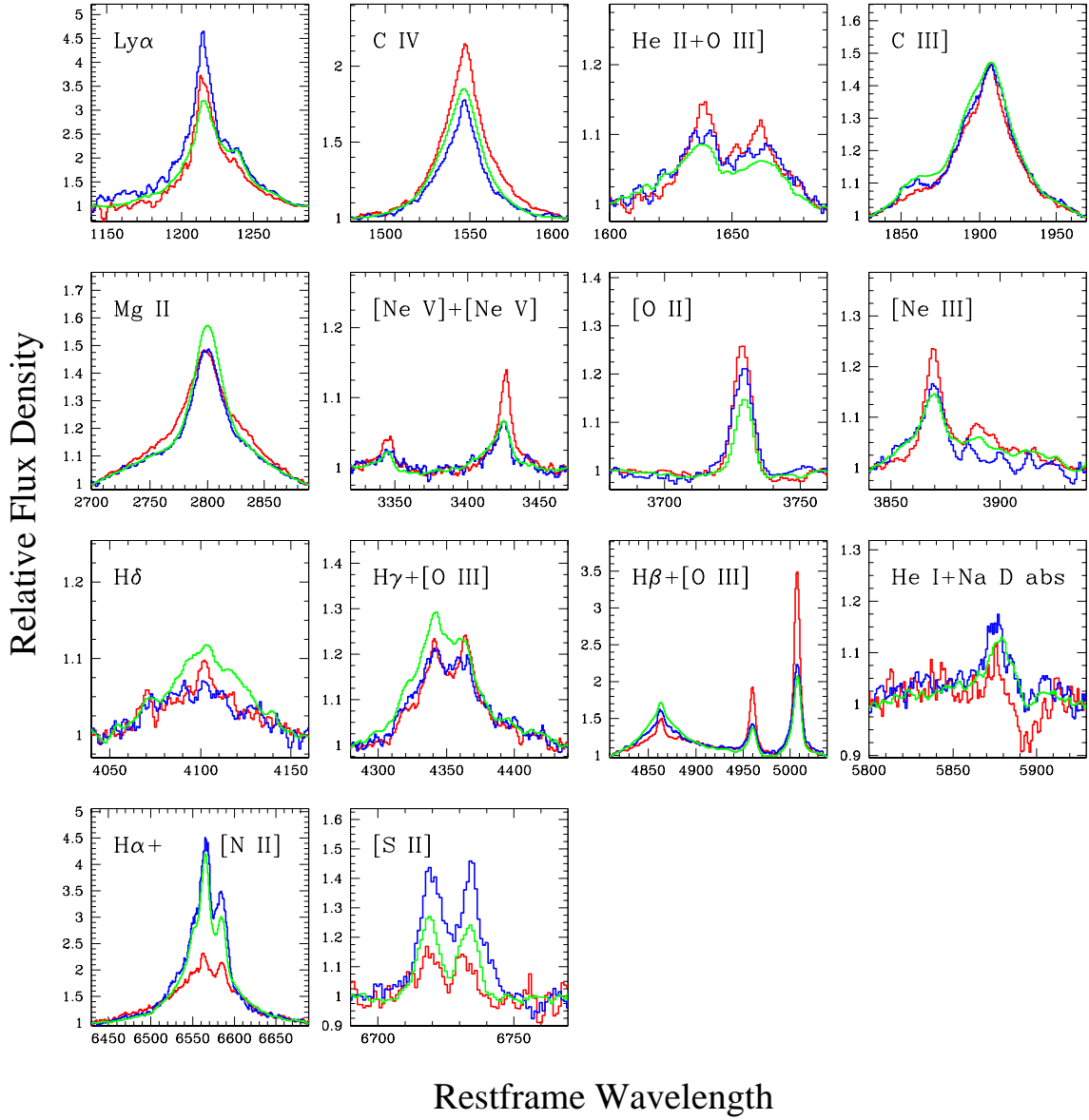}
\caption{Composite spectra of the three comparison samples, centered around
emission line regions. The histograms are color-codes as follows: green is for
the radio quiet (RQ) quasar population, the blue for the core dominated (CD)
sample, and red represents the lobe dominated FR2 quasars. All the spectra have
been normalized to the continuum flux levels at the left and right parts of each
panel.}
\label{specSlide}
\end{figure*}


In order to study differences in line emission, small differences in spectral
slope have to be removed.  This is achieved by first normalizing each spectrum
to the continuum flux just shortward of the emission line in question. Then, by
fitting a powerlaw to the local continuum, each emission line spectrum can be
``rectified'' to a slope of unity (i.e., making sure both the left and right
sides of the zoomed-in spectrum are set to unity). A similar approach has been
employed by \citet[][see their Figs.~8 and 9]{richards03}.

The results are plotted in Fig.~\ref{specSlide}, zoomed in around prominent
emission lines. The panels are arranged in order of increasing restframe
wavelength. A few key observations can be made. The first, and most striking
one, is that FR2 quasars tend to have stronger moderate-to-high ionization
emission lines in their spectrum than either the CD and RQ samples. This can be
seen especially for the \ion{C}{4}, [\ion{Ne}{5}], [\ion{Ne}{3}], and
[\ion{O}{3}] emission lines. The inverse appears to be the case for the Balmer
lines: the FR2 sources have significantly fainter Balmer lines than either the
CD or RQ samples. Notice, for instance, the H$\delta$, H$\gamma$, H$\beta$, and
H$\alpha$ sequence in Fig.~\ref{specSlide}.  Other lines, like \ion{Mg}{2} and
\ion{C}{3}], do not seem to differ among our 3 samples.

\begin{figure}[tb]
\epsscale{1.0}
\plotone{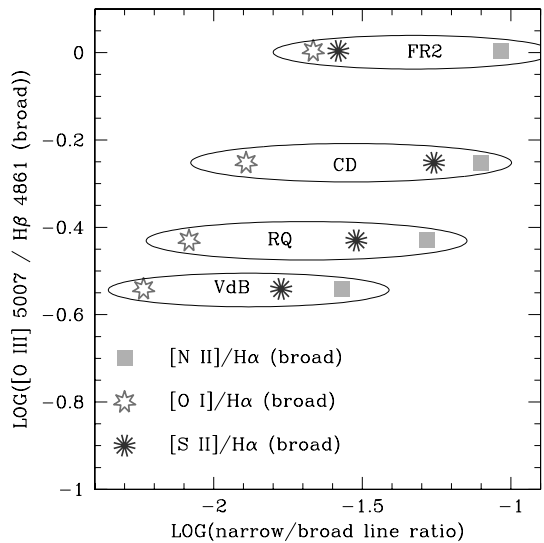}
\caption{Relative importance of broad versus narrow emission lines in our 3
subsamples. For comparison, we included ratios taken from Vanden Berk et
al. (2001), based on a sample of $\sim 2000$ quasars. Datapoints in the lower
left corner can be considered dominated by the broad line component, whereas a
point in the upper right has a more substantial narrow line contribution. The
trend for the FR2 sample to be the one that is most dominated by line emission
is apparent, and consistent for various ratios (as indicated).}
\label{lineDiag}
\end{figure}


Measured line widths, line centers and fluxes for the most prominent emission
lines are listed in Table~6. Since a lot of the lines have shapes that are quite
different from the Gaussian form, we have fitted the profiles with the more
general form $F(x)=c e^{-0.5(x/\sigma)^n}$, with $c$ a normalization constant,
and $n$ a free parameter. Note that for a Gaussian, $n=2$. The FWHM of the
profile can be obtained directly from the values of $n$ and $\sigma$:
$\mbox{FWHM}= 2(2ln2)^{1/n} \sigma$. Allowing values of $n<2$ results in better
fits for lines with broad wings (e.g., \ion{C}{3}$]$ in Fig.~\ref{specSlide}).
Typically the difference in equivalent width (EW) as fitted by the function and
the actual measured value is less than 1\%. The fluxes in Table~6 have been
derived from the measured EW values, multiplied by the continuum level at the
center of the line (as determined by a powerlaw fit, see Fig.~\ref{compView}).
Since all composite spectra have been normalized to a fiducial value of 1.00 at
4000\AA\, the fluxes are relative to this 4000\AA\ continuum value, and can be
compared across the three samples (columns 6 and 11 in Table~6). In addition, we
have normalized these fluxes by the value of the Ly$\alpha$ flux for each
subsample. This effectively takes out the slight spectral slope dependency, and
allows for an easier comparison to the values of \citet[][their Table
2]{vandenberk01}.

The differences between the various species of emission lines among the 3
subsamples, as illustrated in Fig.~\ref{specSlide}, are corroborated by their
line fluxes and ratios. Even though we cannot use emission line ratios (like
$[$\ion{O}{1}$]$ / H$\alpha$, see Osterbrock 1989) to determine whether we are
dealing with AGN or \ion{H}{2}-region dominated emission regimes (due to the
fact that the broad and narrow lines do not originate from the same region), we
can still discern trends between the subsamples in the relative importance of
broad vs. narrow line emission. This is illustrated by Fig.~\ref{lineDiag}, in
which we have plotted various ratios of narrow and broad emission lines (based
on fluxes listed in Table~6). The narrow lines are normalized on the $x$-axis by
the broad-line H$\alpha$ flux, and on the $y$-axis by the broad-line component
(listed separately in Table~6) of the H$\beta$ line. It is clear from this plot
that, as one progresses from RQ, CD, to FR2 sample, the relative importance of
various narrow lines increases. The offset between the RQ and Vanden Berk
samples (which in principle should coincide) is in part due to the presence of a
narrow-line component in their H$\beta$ fluxes (lowering the points along the
$y$-axis), and a slightly larger flux density in their composite H$\alpha$ line
(moving the points to the left along the $x$-axis). The offset probably serves
best to illustrate the inherent uncertainties in plots like these.

So, in summary, it appears that the FR2 sources tend to have brighter
moderate-to-high ionization lines, while at the same time having much less
prominent Balmer lines, than either the CD or RQ samples. The latter two have
far stronger comparable emission line profiles / fluxes, with the possible
exception of the higher Balmer lines and [\ion{S}{2}].

Radio sources are known to interact with their ambient media, especially in the
earlier stages of radio source evolution where the structure is confined to
within the host galaxy. In these compact stages, copious amounts of
line-emission are induced at the interfaces of the radio plasma and ambient
medium \citep[e.g.,][]{bicknell97,devries99,axon99}.  Other types of radio
activity related spectral signatures are enhanced star-formation induced by the
powerful radio jet \citep[e.g.,][]{vanbreugel85,mccarthy87,rees89}, scattered
nuclear UV light off the wall of the area ``excavated'' by the radio structure
\citep[e.g.,][]{diSerego89, dey96}, or more generally, direct photo-ionization
of the ambient gas by the AGN along radiation cones coinciding with the radio
symmetry axis \citep[e.g.,][]{neeser97}. The last three scenarios are more
long-lived (i.e., the resulting stars will be around for a while), whereas the
shock-ionization of the line emission gas is an in-situ event, and will only
last as long as the radio source is there to shock the gas ($< 10^6$ years).

It therefore appears reasonable to guess that in the case of the FR2 quasars,
such an ongoing interaction between the radio structure and its ambient medium
is producing the excess flux in the narrow lines. Indeed, shock precursor clouds
are found to be particularly bright in high-ionization lines like
$[$\ion{O}{3}$]$ compared to H$\alpha$ \citep[e.g.,][]{sutherland93,dopita96}.
Since the optical spectrum is taken {\it at} the quasar position, and not at the
radio {\it lobe} position, we are obviously dealing with interactions between
the gaseous medium and the radio core (whether we detected one or not).

The other sample of quasars associated with radio activity, the core-dominated
(CD) sample, has optical spectral properties which do not differ significantly
from radio-quiet quasars. 

\section{Summary and Conclusions}

We have combined a sample of 44\,984 SDSS quasars with the FIRST radio
survey. Instead of comparing optical and radio positions for the quasars
directly to within a small radius (say, 3\arcsec), we matched the quasar
position to its complete radio {\it environment} within 450\arcsec. This way, we
are able to characterize the radio morphological make-up of what is essentially
an optically selected quasar sample, regardless of whether the quasar (nucleus)
itself has been detected in the radio.

The results can be separated into ones that pertain to the quasar population as
a whole, and those that only concern FR2 sources. For the former category we
list: 1) only a small fraction of the quasars have radio emission associated
with the core itself ($\sim 11\%$ at the 0.75 mJy level); 2) FR2 quasars are
even rarer, only 1.7\% of the general population is associated with a double
lobed radio source; 3) of these, about three-quarter have a detected core; 4)
roughly half of the FR2 quasars have bends larger than 20 degrees from linear,
indicating either interactions of the radio plasma with the ICM or IGM; and 5)
no evidence for correlations with redshift among our FR2 quasars was found:
radio lobe flux densities and radio source diameters of the quasars have similar
distributions at low and high redshifts.

To investigate more detailed source related properties, we used an actual sample
of 422 FR2 quasars and two comparison samples of radio quiet and non-FR2 radio
loud quasars. These three samples are matched in their redshift distributions,
and for each we constructed an optical composite spectrum using SDSS
spectroscopic data. Based on these spectra we conclude that the FR2 quasars have
stronger high-ionization emission lines compared to both the radio quiet and
non-FR2 radio loud sources. This may be due to higher levels of shock ionization
of the ambient gas, as induced by the expanding radio source in FR2 quasars.

\acknowledgments

We like to thank the referee for comments that helped improve the paper.  WDVs
work was performed under the auspices of the U.S. Department of Energy, National
Nuclear Security Administration by the University of California, Lawrence
Livermore National Laboratory under contract No.  W-7405-Eng-48.  The authors
also acknowledge support from the National Radio Astronomy Observatory, the
National Science Foundations (grant AST 00-98355), the Space Telescope Science
Institute, and Microsoft.

\begin{deluxetable}{lrlrlr}
\tablewidth{13cm}
\tablenum{1}
\tablecaption{Radio Core Matching}
\label{corematching}
\tablehead{
  \colhead{} & \colhead{Number} & \colhead{($5\sigma$ matches)\tablenotemark{c}} & \colhead{($3\sigma$ matches)\tablenotemark{c}}
}
\startdata
SDSS Optically Selected\tablenotemark{a} & 34\,147 & 2430 (7.1\%) & 3458 (10.1\%) \\
SDSS Otherwise Selected\tablenotemark{b} & 10\,837 & 1694 (15.6\%) & 1855 (17.1\%) \\
\tableline
Total                   & 44\,984 & 4124 (9.2\%) & 5313 (11.8\%)   \\
\enddata
\tablenotetext{a}{sources which have one (or more) of the following flags set:
QSO\_CAP, QSO\_SKIRT, QSO\_HIZ (as defined in Richards et al. 2002).}
\tablenotetext{b}{sources which do not have any of the flags set mentioned above.}
\tablenotetext{c}{matches to radio sources brighter than either a
5$\sigma$ ($\approx1.0$ mJy), or a 3$\sigma$ ($\approx0.70$ mJy) detection limit.}
\end{deluxetable}

\begin{deluxetable}{lrr}
\tablewidth{110mm}
\tablenum{2}
\tablecaption{Relative Makeup of Quasar Sample}
\label{relmakeup}
\tablehead{\colhead{Total number of quasars} & \colhead{44\,984} & \colhead{Fraction}}
\startdata
Quasars with detected cores (3$\sigma$ limit) & 5\,313 & 11.8\% \\
Non-FR2 quasars with detected cores & 4\,766 & 10.6\% \\
FR2 quasars & 749 & 1.7\% \\
FR2 quasars with detected core & 547 & 73\% \\
FR2 quasars without core & 202 & 27\% \\
\enddata
\end{deluxetable}

\begin{deluxetable}{lrrrr}
\tablewidth{106mm}
\tablenum{3}
\tablecaption{Redshift Matched Subsamples - Properties}
\label{sampleprop}
\tablehead{\colhead{} & \colhead{FR2 sample} & \colhead{CD sample} & \colhead{RQ sample} }
\startdata
Sample Size & 422 & 422 & 6330 \\
\tableline
$u$-band & 19.41$\pm$0.05  & 19.69$\pm$0.06  & 19.61$\pm$0.01\\
$g$-band & 19.11$\pm$0.11  & 19.29$\pm$0.05  & 19.33$\pm$0.01\\
$r$-band & 18.88$\pm$0.04  & 18.98$\pm$0.05  & 19.14$\pm$0.01\\
$i$-band & 18.74$\pm$0.04  & 18.78$\pm$0.05  & 19.01$\pm$0.01\\
$z$-band & 18.64$\pm$0.04  & 18.68$\pm$0.05  & 18.94$\pm$0.01\\
\tableline
Redshift & 1.196$\pm$0.026 & 1.206$\pm$0.027 & 1.209$\pm$0.007\\
\enddata
\end{deluxetable}

\begin{deluxetable}{lllll}
\tabletypesize{\scriptsize}
\tablewidth{180mm}
\tablenum{4}
\tablecaption{FR2 Quasar Sample}
\label{actSample}
\tablehead{ \multicolumn{5}{c}{Optical Positions (J2000)} }
\startdata
00 03 45.23 $-$11 08 18.7 & 00 11 38.44 $-$10 44 58.2 & 00 43 23.43 $-$00 15 52.6 & 00 44 13.73 \phantom{$-$}00 51 41.0 & 00 51 15.12 $-$09 02 08.5\\ 
00 55 08.55 $-$10 52 06.2 & 00 59 29.12 $-$10 17 35.0 & 01 03 29.43 \phantom{$-$}00 40 55.1 & 01 17 49.91 $-$09 05 54.6 & 01 48 47.61 $-$08 19 36.3\\ 
02 35 30.71 $-$07 05 04.5 & 02 40 55.56 \phantom{$-$}00 12 01.2 & 02 48 02.56 $-$07 19 15.8 & 02 50 48.67 \phantom{$-$}00 02 07.6 & 02 55 12.17 $-$00 52 24.5\\ 
03 03 35.76 \phantom{$-$}00 41 45.0 & 03 04 58.97 \phantom{$-$}00 02 35.7 & 03 12 26.12 $-$00 37 08.9 & 03 13 18.66 \phantom{$-$}00 36 23.8 & 07 34 28.88 \phantom{$-$}32 50 58.4\\ 
07 41 00.60 \phantom{$-$}39 55 59.5 & 07 41 25.22 \phantom{$-$}33 33 20.1 & 07 42 42.19 \phantom{$-$}37 44 02.1 & 07 44 51.37 \phantom{$-$}29 20 06.0 & 07 45 41.67 \phantom{$-$}31 42 56.7\\ 
07 48 32.24 \phantom{$-$}26 34 38.9 & 07 49 37.69 \phantom{$-$}38 04 43.4 & 07 52 05.91 \phantom{$-$}28 02 10.8 & 07 52 06.74 \phantom{$-$}24 51 18.8 & 07 52 28.56 \phantom{$-$}37 50 53.6\\ 
07 54 04.24 \phantom{$-$}42 58 04.3 & 07 55 37.03 \phantom{$-$}25 42 39.0 & 07 56 25.82 \phantom{$-$}25 09 21.6 & 07 56 43.10 \phantom{$-$}31 02 48.8 & 07 59 58.76 \phantom{$-$}25 04 38.4\\ 
08 00 52.64 \phantom{$-$}41 57 38.8 & 08 01 29.58 \phantom{$-$}46 26 22.8 & 08 01 39.94 \phantom{$-$}31 53 29.8 & 08 02 20.52 \phantom{$-$}30 35 43.0 & 08 02 50.86 \phantom{$-$}33 35 17.8\\ 
08 05 55.67 \phantom{$-$}34 41 32.3 & 08 06 44.41 \phantom{$-$}48 41 49.2 & 08 08 33.37 \phantom{$-$}42 48 36.4 & 08 10 02.06 \phantom{$-$}26 03 40.5 & 08 11 36.90 \phantom{$-$}28 45 03.6\\ 
08 11 37.23 \phantom{$-$}48 31 33.8 & 08 13 18.85 \phantom{$-$}50 12 39.8 & 08 14 09.22 \phantom{$-$}32 37 32.0 & 08 15 40.84 \phantom{$-$}39 54 37.8 & 08 18 38.88 \phantom{$-$}43 17 55.6\\ 
08 20 50.76 \phantom{$-$}40 34 56.6 & 08 21 17.15 \phantom{$-$}48 45 46.3 & 08 21 25.97 \phantom{$-$}51 37 15.6 & 08 21 33.61 \phantom{$-$}47 02 37.3 & 08 22 14.18 \phantom{$-$}30 54 37.8\\ 
08 23 25.27 \phantom{$-$}44 58 50.5 & 08 26 16.35 \phantom{$-$}52 12 09.3 & 08 30 31.93 \phantom{$-$}05 20 06.8 & 08 31 36.99 \phantom{$-$}30 48 26.8 & 08 32 36.34 \phantom{$-$}33 31 54.7\\ 
08 34 44.62 \phantom{$-$}34 04 14.4 & 08 38 40.58 \phantom{$-$}47 34 10.7 & 08 39 23.24 \phantom{$-$}06 09 58.9 & 08 42 52.40 \phantom{$-$}44 34 10.6 & 08 43 52.87 \phantom{$-$}37 42 28.3\\ 
08 47 02.79 \phantom{$-$}01 30 01.4 & 08 49 25.89 \phantom{$-$}56 41 32.2 & 08 50 39.95 \phantom{$-$}54 37 53.4 & 08 51 14.93 \phantom{$-$}01 59 53.2 & 08 52 00.44 \phantom{$-$}02 29 34.5\\ 
08 52 32.83 \phantom{$-$}38 36 56.1 & 08 53 41.19 \phantom{$-$}40 52 21.8 & 09 01 02.73 \phantom{$-$}42 07 46.4 & 09 02 07.20 \phantom{$-$}57 07 37.9 & 09 02 21.53 \phantom{$-$}48 11 36.5\\ 
09 02 29.23 \phantom{$-$}48 39 06.2 & 09 04 36.36 \phantom{$-$}51 07 28.2 & 09 05 01.56 \phantom{$-$}53 39 07.5 & 09 05 13.17 \phantom{$-$}42 24 51.9 & 09 06 00.09 \phantom{$-$}57 47 30.1\\ 
09 07 45.47 \phantom{$-$}38 27 39.1 & 09 08 08.80 \phantom{$-$}00 35 00.5 & 09 08 12.17 \phantom{$-$}51 47 00.8 & 09 08 21.02 \phantom{$-$}04 50 59.5 & 09 10 17.33 \phantom{$-$}37 42 52.1\\ 
09 12 05.16 \phantom{$-$}54 31 41.3 & 09 15 28.77 \phantom{$-$}44 16 32.9 & 09 17 18.68 \phantom{$-$}59 03 40.4 & 09 17 57.43 \phantom{$-$}02 37 34.0 & 09 18 43.09 \phantom{$-$}40 17 09.7\\ 
09 19 21.56 \phantom{$-$}50 48 55.4 & 09 22 25.14 \phantom{$-$}43 07 49.4 & 09 23 08.17 \phantom{$-$}56 14 55.4 & 09 23 13.54 \phantom{$-$}04 34 45.0 & 09 24 54.56 \phantom{$-$}49 18 04.2\\ 
09 28 37.98 \phantom{$-$}60 25 21.0 & 09 28 56.83 \phantom{$-$}07 36 19.0 & 09 31 38.29 \phantom{$-$}03 15 10.1 & 09 35 18.19 \phantom{$-$}02 04 15.5 & 09 35 53.82 \phantom{$-$}05 03 53.2\\ 
09 36 28.68 \phantom{$-$}01 23 29.3 & 09 41 04.01 \phantom{$-$}38 53 51.0 & 09 41 24.74 \phantom{$-$}03 39 16.4 & 09 41 32.54 \phantom{$-$}02 04 32.6 & 09 41 44.82 \phantom{$-$}57 51 23.7\\ 
09 42 03.05 \phantom{$-$}54 05 18.9 & 09 43 58.22 \phantom{$-$}02 26 30.6 & 09 44 38.79 \phantom{$-$}08 51 57.0 & 09 47 40.01 \phantom{$-$}51 54 56.8 & 09 51 26.48 \phantom{$-$}01 46 51.8\\ 
09 52 28.46 \phantom{$-$}06 28 10.5 & 09 52 45.57 \phantom{$-$}00 00 15.5 & 09 55 56.38 \phantom{$-$}06 16 42.5 & 09 58 02.82 \phantom{$-$}44 06 03.7 & 10 00 54.68 \phantom{$-$}53 32 06.0\\ 
10 03 11.56 \phantom{$-$}50 57 05.2 & 10 03 50.71 \phantom{$-$}52 53 52.2 & 10 05 07.08 \phantom{$-$}50 19 29.6 & 10 06 33.64 \phantom{$-$}44 35 00.7 & 10 09 43.56 \phantom{$-$}05 29 53.9\\ 
10 10 27.52 \phantom{$-$}41 32 39.1 & 10 11 35.45 \phantom{$-$}00 57 43.1 & 10 12 54.25 \phantom{$-$}61 36 34.6 & 10 13 28.77 \phantom{$-$}07 56 53.8 & 10 15 41.14 \phantom{$-$}59 44 45.4\\ 
10 16 09.93 \phantom{$-$}45 31 43.2 & 10 16 51.74 $-$00 33 47.0 & 10 21 06.04 \phantom{$-$}45 23 31.9 & 10 22 35.57 \phantom{$-$}45 41 05.5 & 10 23 20.89 $-$01 02 27.5\\ 
10 26 31.97 \phantom{$-$}06 27 33.0 & 10 27 06.26 \phantom{$-$}06 09 59.2 & 10 27 25.96 \phantom{$-$}52 26 37.0 & 10 30 24.95 \phantom{$-$}55 16 22.8 & 10 31 43.51 \phantom{$-$}52 25 35.2\\ 
10 33 51.88 $-$00 24 14.3 & 10 34 18.01 \phantom{$-$}08 36 26.7 & 10 38 38.82 \phantom{$-$}49 47 36.9 & 10 39 36.67 \phantom{$-$}07 14 27.4 & 10 41 12.53 \phantom{$-$}00 45 49.3\\ 
10 42 07.56 \phantom{$-$}50 13 22.0 & 10 43 12.84 $-$00 13 49.1 & 10 45 34.98 $-$00 55 29.5 & 10 45 49.00 \phantom{$-$}53 47 59.8 & 10 47 26.65 \phantom{$-$}01 11 47.5\\ 
10 47 40.47 \phantom{$-$}02 07 57.3 & 10 49 32.22 \phantom{$-$}05 05 31.7 & 10 51 41.17 \phantom{$-$}59 13 05.2 & 10 53 10.16 \phantom{$-$}58 55 32.8 & 10 53 52.87 $-$00 58 52.8\\ 
10 54 54.97 \phantom{$-$}06 14 53.1 & 10 54 57.04 $-$00 45 53.0 & 10 55 00.34 \phantom{$-$}52 02 00.9 & 10 55 17.27 \phantom{$-$}02 05 45.1 & 10 56 54.16 \phantom{$-$}05 17 13.3\\ 
10 57 51.58 \phantom{$-$}02 27 40.9 & 11 05 14.28 \phantom{$-$}01 09 08.2 & 11 07 09.51 \phantom{$-$}05 47 44.8 & 11 07 15.89 \phantom{$-$}05 33 06.7 & 11 07 18.88 \phantom{$-$}10 04 17.7\\ 
11 07 45.78 \phantom{$-$}60 09 13.9 & 11 12 14.45 \phantom{$-$}01 20 48.5 & 11 12 17.48 \phantom{$-$}61 49 26.5 & 11 17 08.86 \phantom{$-$}00 43 10.6 & 11 17 16.60 \phantom{$-$}57 58 11.8\\ 
11 17 40.84 \phantom{$-$}05 28 58.7 & 11 20 48.50 \phantom{$-$}03 32 47.1 & 11 24 11.66 $-$03 37 04.7 & 11 25 06.96 $-$00 16 47.6 & 11 27 13.99 $-$01 11 53.3\\ 
11 30 44.30 \phantom{$-$}01 37 24.6 & 11 32 35.86 $-$01 28 48.7 & 11 32 50.84 $-$00 20 01.0 & 11 33 03.03 \phantom{$-$}00 15 48.9 & 11 33 25.15 \phantom{$-$}55 11 24.4\\ 
11 36 04.92 \phantom{$-$}52 25 58.9 & 11 37 03.09 \phantom{$-$}01 40 06.3 & 11 37 23.46 $-$02 15 08.2 & 11 41 11.62 $-$01 43 06.7 & 11 43 12.04 \phantom{$-$}53 50 48.3\\ 
11 44 33.67 \phantom{$-$}60 15 38.9 & 11 45 10.39 \phantom{$-$}01 10 56.3 & 11 47 34.30 \phantom{$-$}04 30 47.3 & 11 48 47.83 \phantom{$-$}10 54 58.4 & 11 49 18.97 \phantom{$-$}02 39 26.2\\ 
11 51 44.39 \phantom{$-$}52 44 12.6 & 11 54 05.38 \phantom{$-$}56 20 40.9 & 11 54 09.28 \phantom{$-$}02 38 15.0 & 11 56 51.58 \phantom{$-$}05 03 35.7 & 11 58 39.92 \phantom{$-$}62 54 27.9\\ 
11 59 23.69 \phantom{$-$}01 52 24.2 & 11 59 40.97 $-$01 27 08.5 & 12 05 17.33 \phantom{$-$}00 29 56.9 & 12 07 08.02 $-$02 44 44.1 & 12 09 49.98 \phantom{$-$}54 26 31.6\\ 
12 11 28.87 \phantom{$-$}50 52 54.2 & 12 11 54.87 \phantom{$-$}60 44 26.1 & 12 15 29.56 \phantom{$-$}53 35 55.9 & 12 15 41.96 \phantom{$-$}05 19 32.6 & 12 15 57.60 $-$00 07 19.8\\ 
12 17 01.37 \phantom{$-$}10 19 52.9 & 12 18 58.76 \phantom{$-$}01 52 37.9 & 12 20 27.98 \phantom{$-$}09 28 27.3 & 12 20 49.15 \phantom{$-$}58 59 21.6 & 12 21 30.39 $-$02 41 33.0\\ 
12 31 24.11 \phantom{$-$}01 12 07.3 & 12 36 23.80 \phantom{$-$}06 02 08.2 & 12 36 33.12 \phantom{$-$}10 09 28.7 & 12 37 53.86 \phantom{$-$}00 16 37.8 & 12 38 29.96 $-$00 13 27.9\\ 
12 40 00.93 \phantom{$-$}03 40 51.6 & 12 41 16.49 \phantom{$-$}51 41 30.0 & 12 41 39.73 \phantom{$-$}49 34 05.5 & 12 41 57.55 \phantom{$-$}63 32 41.6 & 12 42 16.38 \phantom{$-$}00 47 43.7\\ 
12 44 47.32 \phantom{$-$}59 41 07.4 & 12 45 38.35 \phantom{$-$}55 11 32.7 & 12 47 10.29 \phantom{$-$}55 55 57.3 & 12 47 58.52 \phantom{$-$}62 50 49.3 & 12 51 51.03 \phantom{$-$}49 18 55.0\\ 
12 53 11.26 \phantom{$-$}01 20 27.3 & 12 54 02.16 $-$00 49 31.1 & 12 55 08.10 \phantom{$-$}62 20 50.5 & 12 55 28.30 $-$00 54 42.0 & 12 55 54.74 \phantom{$-$}57 54 25.4\\ 
12 57 03.12 \phantom{$-$}00 24 36.0 & 12 57 29.79 $-$01 32 39.6 & 12 57 37.07 $-$00 32 20.2 & 12 58 24.70 \phantom{$-$}02 08 46.8 & 12 59 45.18 \phantom{$-$}03 17 26.2\\ 
13 02 10.16 \phantom{$-$}05 08 15.2 & 13 05 21.33 \phantom{$-$}49 51 42.3 & 13 08 28.73 \phantom{$-$}50 26 23.2 & 13 08 42.88 \phantom{$-$}02 43 26.7 & 13 09 07.99 \phantom{$-$}52 24 37.3\\ 
13 10 28.51 \phantom{$-$}00 44 08.9 & 13 10 40.56 $-$03 34 12.0 & 13 10 40.74 \phantom{$-$}02 01 27.1 & 13 15 38.72 $-$01 58 46.2 & 13 16 00.79 $-$02 18 19.6\\ 
13 16 14.50 \phantom{$-$}02 39 38.8 & 13 17 26.13 $-$02 31 50.5 & 13 18 27.00 \phantom{$-$}62 00 36.3 & 13 25 09.70 \phantom{$-$}00 49 33.9 & 13 26 07.16 \phantom{$-$}47 54 41.5\\ 
13 26 31.45 \phantom{$-$}47 37 55.9 & 13 26 55.72 \phantom{$-$}02 07 27.4 & 13 27 46.16 \phantom{$-$}48 42 03.0 & 13 29 09.25 \phantom{$-$}48 01 09.7 & 13 32 21.85 \phantom{$-$}53 28 17.4\\ 
13 32 59.17 \phantom{$-$}49 09 46.8 & 13 34 11.70 \phantom{$-$}55 01 25.0 & 13 34 37.49 \phantom{$-$}56 31 47.9 & 13 38 02.81 \phantom{$-$}42 39 57.0 & 13 39 48.45 \phantom{$-$}47 41 17.1\\ 
13 40 42.88 \phantom{$-$}02 03 07.5 & 13 40 48.37 \phantom{$-$}43 33 59.9 & 13 41 34.85 \phantom{$-$}53 44 43.7 & 13 45 45.36 \phantom{$-$}53 32 52.3 & 13 46 17.55 \phantom{$-$}62 20 45.5\\ 
13 47 39.84 \phantom{$-$}62 21 49.6 & 13 50 54.59 \phantom{$-$}05 22 06.5 & 13 52 23.24 \phantom{$-$}48 46 12.2 & 13 53 05.54 \phantom{$-$}04 43 38.7 & 13 54 09.95 $-$01 41 50.7\\ 
13 57 03.83 \phantom{$-$}02 30 07.1 & 13 58 23.52 \phantom{$-$}60 25 07.3 & 13 59 53.80 \phantom{$-$}59 11 03.0 & 13 59 59.08 $-$01 44 54.2 & 14 01 30.69 \phantom{$-$}41 55 15.3\\ 
14 02 14.80 \phantom{$-$}58 17 46.8 & 14 05 18.48 \phantom{$-$}04 34 06.9 & 14 06 56.49 \phantom{$-$}46 17 12.5 & 14 08 32.48 \phantom{$-$}47 38 37.4 & 14 08 32.65 \phantom{$-$}00 31 38.5\\ 
14 10 30.99 \phantom{$-$}61 41 37.0 & 14 10 54.06 \phantom{$-$}58 46 55.4 & 14 11 23.51 \phantom{$-$}00 42 53.0 & 14 12 31.18 \phantom{$-$}54 55 11.5 & 14 15 32.19 \phantom{$-$}40 49 51.6\\ 
14 17 08.16 \phantom{$-$}46 07 05.4 & 14 17 40.03 \phantom{$-$}60 53 23.8 & 14 19 32.59 \phantom{$-$}00 31 20.3 & 14 20 33.26 $-$00 32 33.3 & 14 21 27.77 \phantom{$-$}58 47 57.6\\ 
14 22 35.89 $-$01 52 11.2 & 14 24 14.09 \phantom{$-$}42 14 00.1 & 14 26 06.19 \phantom{$-$}40 24 32.0 & 14 27 45.71 \phantom{$-$}55 04 10.3 & 14 27 46.79 \phantom{$-$}00 28 44.7\\ 
14 28 29.93 \phantom{$-$}44 39 49.7 & 14 30 17.34 \phantom{$-$}52 17 35.1 & 14 32 44.44 $-$00 59 15.1 & 14 33 39.38 \phantom{$-$}50 52 26.9 & 14 34 10.77 $-$01 23 41.7\\ 
14 36 27.37 \phantom{$-$}52 14 00.0 & 14 37 56.93 \phantom{$-$}01 56 38.9 & 14 38 06.25 $-$00 35 34.8 & 14 38 20.93 $-$02 39 53.1 & 14 38 34.54 \phantom{$-$}39 42 60.0\\ 
14 39 39.95 \phantom{$-$}44 28 51.0 & 14 40 01.64 \phantom{$-$}61 01 33.9 & 14 43 34.57 $-$02 19 26.5 & 14 44 14.21 \phantom{$-$}43 36 55.5 & 14 46 36.90 \phantom{$-$}00 46 56.5\\ 
14 47 07.41 \phantom{$-$}52 03 40.1 & 14 48 14.94 \phantom{$-$}57 41 45.7 & 14 50 49.93 \phantom{$-$}00 01 44.4 & 14 51 34.61 \phantom{$-$}01 59 37.0 & 14 52 40.92 \phantom{$-$}43 08 14.3\\ 
14 52 47.37 \phantom{$-$}47 35 29.2 & 15 00 07.27 \phantom{$-$}56 36 00.8 & 15 00 31.81 \phantom{$-$}48 36 46.8 & 15 01 21.97 \phantom{$-$}01 44 01.2 & 15 04 05.11 \phantom{$-$}46 28 51.3\\ 
15 04 31.30 \phantom{$-$}47 41 51.3 & 15 05 45.05 \phantom{$-$}41 17 26.7 & 15 08 24.74 \phantom{$-$}56 04 23.3 & 15 08 35.94 \phantom{$-$}60 32 58.6 & 15 09 40.59 \phantom{$-$}60 38 21.3\\ 
15 11 09.22 \phantom{$-$}56 50 51.8 & 15 12 15.74 \phantom{$-$}02 03 17.0 & 15 13 25.62 \phantom{$-$}42 00 16.8 & 15 13 34.31 \phantom{$-$}46 12 57.7 & 15 13 56.16 \phantom{$-$}04 20 55.8\\ 
15 14 15.96 \phantom{$-$}57 49 04.0 & 15 15 03.23 \phantom{$-$}61 35 20.1 & 15 19 36.72 \phantom{$-$}53 42 55.5 & 15 20 21.04 \phantom{$-$}02 53 12.1 & 15 20 25.64 \phantom{$-$}59 39 55.4\\ 
15 23 46.79 $-$02 22 39.5 & 15 25 05.36 \phantom{$-$}53 06 22.9 & 15 26 03.80 \phantom{$-$}58 25 25.1 & 15 28 38.40 \phantom{$-$}48 47 40.7 & 15 29 24.65 $-$01 53 43.5\\ 
15 31 07.41 \phantom{$-$}58 44 09.9 & 15 31 28.69 \phantom{$-$}03 38 02.3 & 15 43 16.52 \phantom{$-$}41 36 09.5 & 15 44 30.48 \phantom{$-$}41 20 14.0 & 15 44 39.71 \phantom{$-$}44 40 51.0\\ 
15 52 21.45 \phantom{$-$}00 12 56.9 & 15 54 03.09 \phantom{$-$}32 33 32.6 & 15 57 52.76 \phantom{$-$}02 53 27.9 & 16 03 26.34 \phantom{$-$}49 30 44.3 & 16 05 56.55 \phantom{$-$}26 59 44.2\\ 
16 08 00.02 \phantom{$-$}38 15 30.7 & 16 08 46.76 \phantom{$-$}37 48 50.7 & 16 13 42.98 \phantom{$-$}39 07 32.8 & 16 13 51.34 \phantom{$-$}37 42 58.7 & 16 22 29.93 \phantom{$-$}35 31 25.4\\ 
16 23 36.45 \phantom{$-$}34 19 46.4 & 16 24 21.99 \phantom{$-$}39 24 40.9 & 16 25 13.80 \phantom{$-$}40 58 51.0 & 16 28 51.27 \phantom{$-$}45 52 18.5 & 16 29 17.79 \phantom{$-$}44 34 52.4\\ 
16 29 57.80 \phantom{$-$}42 30 51.4 & 16 30 46.21 \phantom{$-$}36 13 06.0 & 16 35 22.86 \phantom{$-$}39 44 37.8 & 16 35 24.18 \phantom{$-$}31 10 00.4 & 16 36 09.24 \phantom{$-$}26 23 09.1\\ 
16 36 24.31 \phantom{$-$}47 15 35.8 & 16 36 40.56 \phantom{$-$}46 47 07.2 & 16 36 54.41 \phantom{$-$}32 20 06.5 & 16 37 02.21 \phantom{$-$}41 30 22.2 & 16 38 56.54 \phantom{$-$}43 35 12.6\\ 
16 39 55.99 \phantom{$-$}47 05 23.6 & 16 40 54.16 \phantom{$-$}31 43 29.9 & 16 44 52.58 \phantom{$-$}37 30 09.3 & 16 45 44.69 \phantom{$-$}37 55 26.1 & 16 46 34.71 \phantom{$-$}35 03 17.6\\ 
16 50 27.67 \phantom{$-$}36 22 56.5 & 16 58 19.55 \phantom{$-$}62 38 23.2 & 16 58 20.35 \phantom{$-$}37 33 15.6 & 16 59 12.68 \phantom{$-$}40 03 59.1 & 16 59 43.08 \phantom{$-$}37 54 22.7\\ 
17 01 23.98 \phantom{$-$}38 51 37.1 & 17 03 35.00 \phantom{$-$}39 17 35.6 & 17 04 04.50 \phantom{$-$}38 54 30.7 & 17 04 49.89 \phantom{$-$}31 17 28.2 & 17 05 18.52 \phantom{$-$}35 33 52.3\\ 
17 06 48.06 \phantom{$-$}32 14 22.9 & 17 06 51.42 \phantom{$-$}38 16 45.3 & 17 08 01.25 \phantom{$-$}33 46 46.3 & 17 13 28.81 \phantom{$-$}30 59 07.8 & 17 14 30.10 \phantom{$-$}61 57 46.6\\ 
17 15 54.63 \phantom{$-$}28 44 49.9 & 17 20 51.15 \phantom{$-$}62 09 44.3 & 17 21 58.61 \phantom{$-$}55 47 07.5 & 21 07 54.98 $-$06 25 19.1 & 21 08 18.46 $-$06 17 40.5\\ 
21 27 15.34 $-$06 20 41.7 & 21 30 04.76 $-$01 02 44.4 & 21 35 13.10 $-$00 52 43.9 & 21 41 11.89 $-$06 39 30.3 & 21 44 32.75 $-$07 54 42.8\\ 
21 49 37.38 $-$06 53 12.9 & 21 52 13.51 $-$07 42 24.9 & 21 59 34.46 $-$09 10 22.2 & 22 07 52.53 $-$00 17 35.8 & 22 14 09.97 \phantom{$-$}00 52 27.0\\ 
22 14 26.04 $-$08 16 37.7 & 22 27 20.56 $-$00 01 59.3 & 22 27 29.06 \phantom{$-$}00 05 22.0 & 22 29 12.25 $-$09 42 18.9 & 22 55 10.38 $-$09 07 55.1\\ 
23 05 45.67 $-$00 36 08.6 & 23 12 12.16 $-$09 19 28.7 & 23 20 20.78 $-$00 20 10.5 & 23 21 33.77 $-$01 06 46.0 & 23 35 34.68 $-$09 27 39.2\\ 
23 41 42.32 \phantom{$-$}00 33 12.6 & 23 45 40.45 $-$09 36 10.3 & 23 47 24.71 \phantom{$-$}00 52 47.0 & 23 50 06.26 $-$00 09 33.4 & 23 50 26.40 $-$10 19 58.1\\ 
23 51 56.13 $-$01 09 13.4 & 23 53 21.04 $-$08 59 30.6 \\
\enddata
\end{deluxetable}

\begin{deluxetable}{lrlrlr}
\tablenum{5}
\tablewidth{158mm}
\tablecaption{FR2 Sample Properties}
\label{FR2sampleprop}
\tablehead{\colhead{Armlength Ratio $Q$} & \colhead{} & \colhead{Opening Angle $\Psi$} & \colhead{(\arcdeg)} & 
\colhead{Lobe Flux Ratio $F$} & \colhead{} }
\startdata
$Q_{\rm mean}$          & 1.90 & $\Psi_{\rm mean}$          & 171.3 & $F_{\rm mean}$         & 1.476\\
$Q_{\rm mean,low z}$    & 1.92 & $\Psi_{\rm mean,low z}$    & 171.7 & $F_{\rm mean,low z}$   & 1.411\\
$Q_{\rm mean,high z}$   & 1.88 & $\Psi_{\rm mean,high z}$   & 170.9 & $F_{\rm mean,high z}$  & 1.540\\
\tableline
$Q_{\rm median}$        & 1.48 & $\Psi_{\rm median}$        & 173.2 & $F_{\rm median}$       & 0.859\\
$Q_{\rm median,low z}$  & 1.52 & $\Psi_{\rm median,low z}$  & 173.6 & $F_{\rm median,low z}$ & 0.940\\
$Q_{\rm median,high z}$ & 1.44 & $\Psi_{\rm median,high z}$ & 172.7 & $F_{\rm median,high z}$& 0.813\\
\enddata

\tablecomments{The $F$ distribution is not symmetric around $F=1$, and therefore
the median and mean values differ significantly.}
\end{deluxetable}

\clearpage
\renewcommand{\arraystretch}{.6}
\setlength{\hoffset}{-15mm}
\setlength{\voffset}{20mm}
\setlength{\textheight}{250mm}

\begin{deluxetable}{lrcrrrcrcrrrcrcrrr}
\rotate
\tabletypesize{\scriptsize}
\tablewidth{231mm}
\tablenum{6}
\tablecaption{Composite Spectrum Emission Line Properties}
\label{lineprop}
\tablehead{
\\
\colhead{} & 
\multicolumn{5}{c}{FR2 sample} & \colhead{} & 
\multicolumn{5}{c}{CD sample}  & \colhead{} & 
\multicolumn{5}{c}{RQ sample} 
\\
\cline{2-6} \cline{8-12} \cline{14-18}
\\
\colhead{Line} & 
\colhead{$\lambda_{\rm cent}$} & \colhead{FWHM} & \colhead{EW} & \colhead{Flux\tablenotemark{a}} & \colhead{Flux\tablenotemark{b}} & \colhead{\phantom{.}} &
\colhead{$\lambda_{\rm cent}$} & \colhead{FWHM} & \colhead{EW} & \colhead{Flux\tablenotemark{a}} & \colhead{Flux\tablenotemark{b}} & \colhead{\phantom{.}} &
\colhead{$\lambda_{\rm cent}$} & \colhead{FWHM} & \colhead{EW} & \colhead{Flux\tablenotemark{a}} & \colhead{Flux\tablenotemark{b}}
\\
\colhead{} & 
\colhead{(\AA)} & \colhead{(\AA)} & \colhead{(\AA)} & \colhead{/F$_{\rm Ly\alpha}$} & \colhead{/F$_{\rm 4000}$} & \colhead{} &
\colhead{(\AA)} & \colhead{(\AA)} & \colhead{(\AA)} & \colhead{/F$_{\rm Ly\alpha}$} & \colhead{/F$_{\rm 4000}$} & \colhead{} &
\colhead{(\AA)} & \colhead{(\AA)} & \colhead{(\AA)} & \colhead{/F$_{\rm Ly\alpha}$} & \colhead{/F$_{\rm 4000}$} }
\startdata
Ly$\alpha$        & 1216.5 & 15.4 & 74.50 & 100.0 & 406.7 & & 1215.7 & 13.2 & 81.20 & 100.0 & 347.0 & & 1218.6 & 29.7 & 75.22 & 100.0 & 428.9  \\
C {\rm IV}        & 1546.8 & 21.0 & 35.81 & 35.55 & 144.6 & & 1545.9 & 19.7 & 22.44 & 21.10 & 73.21 & & 1545.5 & 24.1 & 27.40 & 26.94 & 115.6  \\
He {\rm II}       & 1637.9 & 11.2 &  2.31 & 2.134 & 8.679 & & 1637.3 & 21.3 &  2.54 & 2.239 & 7.769 & & 1636.5 & 21.6 &  2.47 & 2.259 & 9.688  \\
O {\rm III}$]$    & 1661.4 & 13.4 &  1.65 & 1.497 & 6.089 & & 1664.4 & 19.6 &  1.22 & 1.056 & 3.663 & & 1664.2 & 17.0 &  0.75 & 0.671 & 2.880  \\
Al {\rm III}      & 1858.2 & 35.8 &  1.50 & 1.183 & 4.810 & & 1858.1 & 15.1 &  1.85 & 1.415 & 4.909 & & 1857.7 & 26.9 &  2.07 & 1.612 & 6.913  \\
C {\rm III}$]$    & 1905.6 & 29.8 & 18.45 & 14.09 & 57.32 & & 1905.9 & 33.5 & 19.54 & 14.52 & 50.39 & & 1905.0 & 37.0 & 20.99 & 15.83 & 67.89  \\
Mg {\rm II} (b)   & 2797.8 &106.0 & 29.04 & 13.69 & 55.69 & & 2799.3 &104.7 & 24.95 & 12.04 & 41.79 & & 2797.9 &126.6 & 22.51 & 10.42 & 44.70  \\
Mg {\rm II} (n)   & 2797.8 & 23.2 &  8.81 & 4.154 & 16.90 & & 2799.8 & 24.2 &  7.75 & 3.740 & 12.98 & & 2800.4 & 25.9 & 12.52 & 5.790 & 24.84  \\
$[$Ne {\rm V}$]$  & 3344.8 &  9.0 &  0.60 & 0.226 & 0.920 & & 3345.1 &  9.9 &  0.42 & 0.166 & 0.576 & & 3344.1 &  8.4 &  0.33 & 0.121 & 0.523 \\
$[$Ne {\rm V}$]$  & 3426.3 &  6.8 &  1.87 & 0.684 & 2.780 & & 3424.4 &  7.9 &  1.02 & 0.393 & 1.362 & & 3423.3 & 13.1 &  1.33 & 0.477 & 2.045 \\
$[$O {\rm II}$]$  & 3728.6 &  8.5 &  2.27 & 0.746 & 3.035 & & 3728.7 &  9.4 &  2.11 & 0.738 & 2.561 & & 3729.2 &  8.2 &  1.29 & 0.415 & 1.779 \\
$[$Ne {\rm III}$]$& 3869.2 &  7.6 &  1.49 & 0.468 & 1.902 & & 3868.7 &  9.0 &  1.88 & 0.631 & 2.189 & & 3868.3 & 10.1 &  1.49 & 0.457 & 1.962  \\
H$_\delta$        & 4101.0 & 29.3 &  3.73 & 1.088 & 4.425 & & 4098.0 & 62.0 &  3.50 & 1.101 & 3.821 & & 4102.5 & 46.4 &  5.58 & 1.590 & 6.818  \\
H$_\gamma$        & 4340.6 & 17.2 &  9.02 & 2.450 & 9.964 & & 4339.8 & 33.7 &  8.12 & 2.395 & 8.311 & & 4340.9 & 32.8 & 12.46 & 3.304 & 14.17  \\
$[$O {\rm III}$]$ & 4364.4 &  8.0 &  5.44 & 1.468 & 5.968 & & 4364.3 & 26.7 &  6.98 & 2.046 & 7.100 & & 4363.6 & 14.2 &  6.13 & 1.615 & 6.926 \\
H$_\beta$ (b)     & 4862.1 & 66.1 & 20.20 & 4.758 & 19.35 & & 4862.4 & 43.3 & 23.52 & 6.107 & 21.19 & & 4862.5 & 40.0 & 32.12 & 7.375 & 31.63 \\
H$_\beta$ (n)     & 4862.1 &  7.8 &  2.30 & 0.542 & 2.203 & & 4862.4 &  6.9 &  1.62 & 0.421 & 1.459 & & 4862.5 &  7.1 &  1.67 & 0.384 & 1.645 \\
$[$O {\rm III}$]$ & 4959.7 &  7.5 &  6.62 & 1.521 & 6.185 & & 4959.6 & 10.2 &  3.76 & 0.955 & 3.313 & & 4959.6 &  9.7 &  3.53 & 0.791 & 3.390  \\
$[$O {\rm III}$]$ & 5007.6 &  7.2 & 21.17 & 4.805 & 19.54 & & 5007.5 &  9.2 & 13.61 & 3.419 & 11.86 & & 5007.7 &  8.9 & 12.37 & 2.736 & 11.74  \\
$[$O {\rm I}$]$   & 6301.5 & 17.3 &  1.95 & 0.332 & 1.349 & & 6301.0 & 14.0 &  1.96 & 0.380 & 1.320 & & 6302.9 &  9.7 &  1.15 & 0.190 & 0.815 \\
H$_\alpha$        & 6561.3 & 58.4 & 95.02 & 15.36 & 62.47 & & 6564.9 & 22.8 &159.52 & 29.56 & 102.6 & & 6564.6 & 20.6 &146.55 & 22.99 & 98.62  \\
$[$N {\rm II}$]$  & 6584.3 & 11.1 &  8.87 & 1.428 & 5.806 & & 6586.1 & 12.2 & 12.71 & 2.347 & 8.144 & & 6585.9 &  8.6 &  7.66 & 1.197 & 5.133  \\
$[$S {\rm II}$]$  & 6718.7 &  8.2 &  1.64 & 0.257 & 1.047 & & 6719.5 &  8.8 &  4.59 & 0.829 & 2.875 & & 6718.8 &  8.2 &  2.55 & 0.388 & 1.666  \\
$[$S {\rm II}$]$  & 6732.0 &  8.1 &  0.93 & 0.146 & 0.592 & & 6734.2 &  6.7 &  4.45 & 0.802 & 2.781 & & 6733.3 &  8.8 &  2.01 & 0.305 & 1.310  \\
\enddata

\tablecomments{Some of the line-centers are offset from their laboratory
wavelengths, and / or display systematic offsets relative to other emission
lines. This can be explained either in terms of possible gas flows in the Broad
Line Region (BLR) relative to the Narrow Line Region (NLR), dust attenuation,
or possibly relativistic effects.  See vanden Berk et al. (2001) for more
discussion.}

\tablenotetext{a}{Relative flux (in percent) to the flux in the Ly$\alpha$ line
of each subsample.}
\tablenotetext{b}{Relative flux in units of the continuum level at 4000\AA,
which has been set to unity for all three samples.}
\end{deluxetable}

\end{document}